\documentclass[12pt]{article}
\usepackage{graphicx, amsmath, amssymb, color, array, pdfpages, pdflscape, fullpage}
\usepackage[numbers, sort&compress]{natbib}

\newcommand{\beq}{\begin{equation}}
\newcommand{\eeq}{\end{equation}}
\newcommand{\s}{\sigma}
\newcommand{\F}{\mathcal{F}}

\begin{document}

\title{Biophysical Fitness Landscapes for Transcription Factor Binding Sites}
\author{Allan Haldane$^1$, Michael Manhart$^1$, and Alexandre V. Morozov$^{1,2}$\footnote{Corresponding author: \texttt{morozov@physics.rutgers.edu}} \\ 
	\small{\emph{$^1$ Department of Physics and Astronomy, Rutgers University, Piscataway, NJ 08854, USA}} \\
	\small{\emph{$^2$ BioMaPS Institute for Quantitative Biology, Rutgers University, Piscataway, NJ 08854, USA}}
	}
\date{}
\maketitle

\begin{abstract}
Evolutionary trajectories and phenotypic states available to cell populations are ultimately dictated by intermolecular interactions between DNA, RNA, proteins, and other molecular species. Here we study how evolution of gene
regulation in a single-cell eukaryote \textit{S. cerevisiae} is affected by the interactions between transcription factors (TFs) and their cognate genomic sites. Our study is informed by high-throughput \textit{in vitro} measurements of TF-DNA binding interactions and by a comprehensive collection of 
genomic binding sites. Using an evolutionary model for monomorphic populations evolving on a fitness landscape, we infer fitness as a function
of TF-DNA binding energy for a collection of 12 yeast TFs, and show that 
the shape of the predicted fitness functions is in broad agreement with a simple
thermodynamic model of two-state TF-DNA binding. However, the effective
temperature of the model is not always equal to the physical temperature,
indicating selection pressures in addition to biophysical constraints
caused by TF-DNA interactions. We find little statistical support for the
fitness landscape in which each position in the binding site evolves
independently, showing that epistasis is common in evolution of gene
regulation. Finally, by correlating TF-DNA binding energies with biological
properties of the sites or the genes they regulate, we are able to rule out several scenarios of site-specific selection, under which binding sites of the same TF would experience a spectrum of selection pressures
depending on their position in the genome.
These findings argue for the
existence of universal fitness landscapes which shape evolution of all
sites for a given TF, and whose properties are determined in part by the
physics of protein-DNA interactions.
\end{abstract}

\vspace{.5cm}
\textbf{Short Title}: Biophysics and Evolution of Gene Regulation


\section*{Author Summary}

Specialized proteins called transcription factors turn genes on and off by binding to short stretches of DNA in their vicinity. Precise gene
regulation is essential for cellular survival and proliferation, and its evolution and maintenance under mutational pressure are central issues in biology.
Here we discuss how evolution of gene regulation is shaped by the need to maintain favorable binding energies between transcription factors and their genomic binding sites. We show that, surprisingly, transcription factor
binding is not affected by the essentiality of the gene it regulates. Rather,
all sites for a given factor appear to evolve under a universal set of constraints, which to a first approximation can be understood in terms
of simple binding thermodynamics.

\section{Introduction}

     A powerful concept in evolution is the fitness landscape: for every possible genotype there is a number, known as the genotypic fitness, that characterizes the evolutionary success of that genotype~\cite{Wright1932}.  Evolutionary success is typically quantified as the probability of surviving to reproduce, number of offspring, growth rate, or a related proxy~\cite{Orr2009, Szendro2013}.  The structure of the fitness landscape is key to understanding the evolutionary fates of populations.

     Most traditional studies of molecular evolution rely on simplified models of fitness landscapes, or reconstruct the landscapes empirically based on limited experimental data~\cite{Szendro2013}. However, fitness landscapes are fundamentally shaped by complex molecular interactions involving DNA, RNA, proteins, and other molecular species present in the cell. Thus we should be able to cast these landscapes in terms of biophysical properties such as binding affinities, molecular stabilites, and degradation rates. The increasing availability of quantitative high-throughput data on molecular interactions in the cell has led to growing efforts aimed at developing models of evolution that explicitly incorporate the underlying biophysics~\cite{Sengupta2002, Gerland2002, Berg2003, Berg2004, Bloom2005, DePristo2005, Bloom2006, Zeldovich2007, Lassig2007, Bloom2007, Bershtein2008, Mustonen2008, Bloom2009, Manhart2013a, Manhart2013b}.  These models combine evolutionary theory with physical models of molecular systems, for example focusing on how protein folding stability or specificity of intermolecular interactions shapes the ensemble of accessible evolutionary pathways and steady-state distributions of biophysical phenotypes.

Evolution of gene regulation is particularly well-suited to this type of analysis. Gene activation and repression are mediated by binding of 
transcription factors (TFs) to their cognate genomic sites.  TF binding sites are short nucleotide sequences, typically 5-25 bp in length, in gene promoters that interact specifically with TF DNA-binding domains~\cite{Ptashne2002}.  In eukaryotes, a given TF can have numerous binding sites in the genome, and many genes are regulated by several TFs~\cite{Ptashne2002,Rhee2011}. Understanding TF-mediated regulation is key to understanding the complex regulatory networks within eukaryotic cells --- one of the main challenges facing molecular biology.  Moreover, the availability of high-throughput datasets on the genomic locations of TF binding sites~\cite{Lee2002, Harbison2004, MacIsaac2006, Chen2010}, and on TF-DNA energetics~\cite{Stormo1998, Berger2006, Foat2006, Fordyce2010} 
make it possible to develop biophysical models of evolution of gene regulation.

     Here we consider evolution of TF binding sites in the yeast \textit{S. cerevisiae}. We study how energetics of protein-DNA interactions affects the structure of the fitness landscape.  We analyze a collection of 25 \textit{S. cerevisiae} TFs for which models of TF binding affinity and specificity were built using high-throughput \textit{in vitro} measurements of TF-DNA interactions~\cite{Fordyce2010}. We focus on 12 TFs for which sufficient data on genomic sites~\cite{Chen2010} is also available.  We use a model of monomorphic populations undergoing consecutive substitutions~\cite{Kimura1983, Sella2005, Lassig2007, Manhart2012} to infer fitness landscapes, as a function of TF binding energy, from observed distributions of TF binding sites in the yeast genome~\cite{Mustonen2008}.  We rationalize these fitness landscapes in terms of a two-state thermodynamic model of TF-DNA binding. Our analysis sheds light on the genome-wide importance of TF-DNA interactions in regulatory site evolution.

	Specifically, we investigate the hypothesis that universal biophysical constraints rather than site-specific selective pressures dominate evolution of regulatory sites. We test the relationship between TF binding energies and various biological properties, such as the essentiality of the corresponding gene~\cite{Winzeler1999}. We find no clear relationship between physical and biological properties of TF sites, which indicates that evolution of site energetics is largely insensitive to site-specific biological functions and is therefore driven by global biophysical constraints.


\section{Biophysical model of TF binding site evolution}


\subsection{Energetics of TF-DNA binding}


     The probability of a binding site to be TF-bound is given by the Fermi-Dirac function of the free energy $E$ of TF-DNA interaction~\cite{Berg1987}:
     
\beq
p_{\text{bound}}(E) = \frac{1}{1 + e^{\beta(E - \mu)}},
\label{eq:prob_bound}
\eeq

\noindent where $\beta$ is the inverse temperature ($\approx 1.7$ (kcal/mol)$^{-1}$ at room temperature) and $\mu$ is the chemical potential, a function of the TF concentration.  The binding energy $E = E(\s)$ of a site is a function of its nucleotide sequence, $\s = (\s_1,\ldots,\s_L)$, where $L$ is the length of the site and $\s_i \in \{\mathsf{A,C,G,T}\}$.
Note that $p_\text{bound}(E) \approx e^{-\beta(E - \mu)}$ if $E \gg \mu$,
resulting in a Boltzmann-like exponential distribution.
In the mean-field approximation, each nucleotide makes an additive contribution to the total energy of the site~\cite{Stormo1998}. These contributions are parameterized by an energy matrix (EM), whose entries $\epsilon^{\s_i}_i$ give the contribution to the total energy from the nucleotide $\s_i$ at position $i$:
     
\beq
E(\s) = \sum_{i=1}^L \epsilon^{\s_i}_i.
\label{eq:seq_energy}
\eeq

\noindent EMs can be readily generalized to more complex models of sequence-dependent energetics, such as those with contributions from dinucleotides, although here we use the additive model.



\subsection{Evolutionary model}

     We consider a population with a locus in the monomorphic limit: mutations in the locus are infrequent enough that each new mutation either fixes or goes extinct before a second mutation arises~\cite{Kimura1983}.  This approximation is valid in the limit $u \ll (L N_e \log N_e)^{-1}$, where $u$ is the mutation rate (probability of mutation per base per generation), $L$ is the number of bases in the locus, and $N_e$ is an effective population size~\cite{Champagnat2006}.  We assume that the locus is unlinked to the rest of the genome by recombination, and thus we can consider its evolution independently.  In evolutionary steady state, the probability that the population has genotype $\s$ at the locus is given by~\cite{Sella2005, Lassig2007, Manhart2012}

\beq
\pi(\s) = \frac{1}{Z} \pi_0(\s) \F(\s)^\nu,
\label{eq:steady_state}
\eeq

\noindent where $\F(\s)$ is the multiplicative fitness (defined so that the total fitness of a set of independently evolving loci is a product of fitnesses of each one), $\pi_0(\s)$ is the neutral distribution of sequences (steady state under no selection), and $Z$ is a normalization constant.  The exponent $\nu$ is a ``scaling'' effective population size which is closely related to the standard variance effective population size $N_e$~\cite{Manhart2012}.  For example, $\nu = 2(N_e-1)$ in the Wright-Fisher model and $\nu = N_e-1$ in the Moran model of
population genetics~\cite{Ewens2004}. Conceptually, both $\nu$ with $N_e$ measure the strength of genetic drift~\cite{Kimura1983}.

     The distribution in Eq.~\ref{eq:steady_state} is applicable to a wide class of population models~\cite{Manhart2012} (see Methods for details).  An analogy with statistical mechanics is suggested by rewriting Eq.~\ref{eq:steady_state} as a Boltzmann distribution:

\beq
\pi(\s) = \frac{1}{Z} \pi_0(\s) e^{\nu \log \F(\s) }.
\eeq

\noindent Here the logarithm of fitness plays the role of a negative Hamiltonian, and the neutral distribution $\pi_0(\s)$ plays the role of entropy.  Typically we expect relatively few sequences with high fitness and many with low fitness; thus mutations drive the population toward lower fitness, while selection favors higher fitness. The balance between these two competing forces depends on the effective population size $\nu$, which controls the strength of random fluctuations and is analogous to inverse temperature in the Boltzmann distribution.


\subsection{Biophysical model of binding site evolution}
\label{sec:biophysical_model}

Since we are primarily interested in the biophysical aspects of binding site evolution, it is more convenient to consider evolution in the space of binding energies by projecting Eq.~\ref{eq:steady_state} via the sequence-energy mapping of Eq.~\ref{eq:seq_energy}:
     
\beq
\pi(E) = \frac{1}{Z} \pi_0(E) \F(E)^\nu.
\label{eq:energy_proj}
\eeq
     
\noindent Here, the binding site fitness $\F(E)$ depends only on the binding energy $E$. We assume that a site contributes fitness 1 to the organism when it is bound, and fitness $f_0 < 1$ otherwise.  Then the fitness contribution, averaged over the bound and unbound states of the TF-DNA
complex, is given by
     
\beq
\F(E) = \frac{1 + f_0 e^{\beta(E - \mu)}}{1 + e^{\beta(E - \mu)}}.
\label{eq:FD_fitness}
\eeq


     An important feature of Eq.~\ref{eq:steady_state} is that we may invert it to obtain the fitness function in terms of the observed steady-state distributions $\pi(\s)$ and $\pi_0(\s)$, or $\pi(E)$ and $\pi_0(E)$ in energy space~\cite{Lassig2007}:
     
\beq
\log\left( \frac{\pi(\s)}{\pi_0(\s)} \right) = \nu \log \F(\s) - \log Z \quad \Longrightarrow \quad \log\left( \frac{\pi(E)}{\pi_0(E)} \right) = \nu \log \F(E) - \log Z.
\label{eq:inference}
\eeq

\noindent Thus given a distribution of evolved binding site sequences $\pi$ and a neutral distribution $\pi_0$, we can use Eq.~\ref{eq:inference} to infer the logarithm of the fitness landscape up to an overall scale and shift.  Moreover, given a specific functional form of $\F(E)$, such as the Fermi-Dirac fitness in Eq.~\ref{eq:FD_fitness}, we can perform a maximum likelihood fit of the observed sequence distribution to infer values of parameters $\beta$, $\mu$, $\nu$, and $f_0$.

     When $1-f_0 \ll 1$, $\F(E)^\nu$ contains an approximate degeneracy in terms of $\nu (1-f_0) \equiv \gamma$, i.e., all fitness functions with constant $\gamma$ are approximately equivalent.  This is a general property of a model where fitness is an average over two possible phenotypes.  Consider a general fitness function

\beq
\F(\s) = p(\s) + f_0 (1 - p(\s)),
\eeq

\noindent where one phenotype has fitness $1$ and occurs with probability $p(\s)$, and the other phenotype has fitness $f_0$ and occurs with probability $1 - p(\s)$.  In the case of binding sites, the phenotypes are TF-bound and TF-unbound, and $p(\s)$ is a Fermi-Dirac function projected from the genotype $\s$ to the energy (Eq.~\ref{eq:prob_bound}).
The steady-state distribution (Eq.~\ref{eq:steady_state}) depends on the quantity $\F(\s)^\nu$, which can be written as:

\beq
\F(\s)^\nu = (1 - \frac{1}{\nu} \gamma (1 - p(\s)))^\nu \approx e^{-\gamma(1 - p(\s))}
\eeq
if $\gamma (1 - p(\s)) \ll \nu$ or, since $0 \leq 1 - p(\s) \leq 1$, if $1-f_0 \ll 1$.  Therefore in this limit, the steady-state distribution $\pi(\s)$ depends only on the parameter $\gamma$ and not on $f_0$ and $\nu$ separately.

     This degeneracy in the steady-state distribution is not surprising in light of the underlying population genetics.  The quantity $1-f_0$ is the selection coefficient $s$ between the two phenotypes of the system, e.g., the bound and unbound states of the TF binding site.  As discussed above, the quantity $\nu$ is an effective population size, which sets the strength $1/\nu$ of genetic drift.  
When $s \ll 1$ and $\nu \gg 1$, steady-state properties of the population depend only on the strength of selection relative to the strength of drift~\cite{Ewens2004, Wakeley2005}, $Ns$, or in our model, $\nu (1-f_0) = \gamma$.
Note that only the absolute magnitude of the selection coefficient $s = 1-f_0$ is required to be small for this degeneracy to hold; the selection strength relative to drift $Ns = \gamma$ may still be large.



\subsection{Selection strength and its dependence on biophysical parameters}

     We now consider how changes to biophysical parameters of the model affect the strength of selection on binding sites.  The selection coefficient for a mutation with small change in energy $\Delta E$ is

\beq
s(E) = \frac{\F(E + \Delta E)}{\F(E)} - 1 \approx \frac{d\log\F}{dE} \Delta E.
\eeq

\noindent Therefore we can characterize local variations in the strength of selection by
considering $\tilde{s}(E) = |d\log\F/dE|$, the per-unit-energy local selection coefficient.
For the Fermi-Dirac landscape, we obtain

\beq
\tilde{s}(E) = \left| \frac{d}{dE} \log \F(E) \right| = \frac{\beta(1-f_0)z}{(1 + z)(1 + f_0z)}, \quad \text{where} \quad z = e^{\beta(E-\mu)}.
\label{eq:sE}
\eeq

\noindent We use the absolute value here since the sign of the selection coefficient is always unambiguous, as the Fermi-Dirac function decreases monotonically with energy.

\begin{figure}[ht!]
\centering\includegraphics[width=\textwidth]{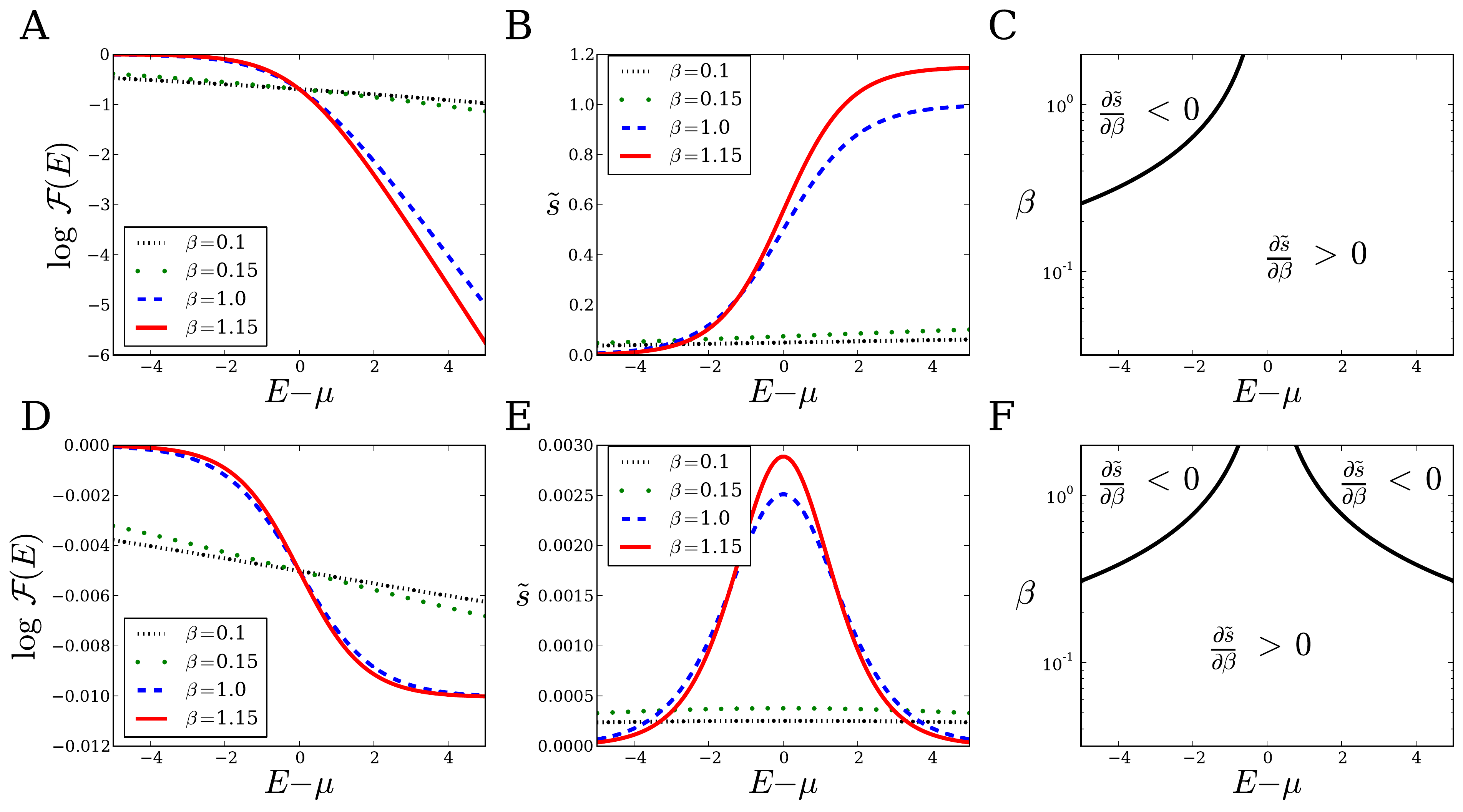}
\caption{
Fitness and selection strength plots as functions of energy $E-\mu$ (measured with respect to chemical potential $\mu$) and
inverse temperature $\beta$. Top row uses $f_0 = 0$; bottom row uses $f_0 = 0.99$.
(A,D) Logarithm of Fermi-Dirac fitness versus energy for several values of $\beta$; note that the high-energy tail looks distinctly different when $f_0$ is nonzero.
(B,E) Per-unit-energy selection strength $\tilde{s}$ versus energy for several values of $\beta$; note that the relative ordering of selection strength curves depends on the value of $E-\mu$.
(C,F) Sign of derivative of selection strength with respect to $\beta$, as a function of $E-\mu$ and $\beta$.  Black boundary in (C) is the curve $\beta(E-\mu) = \log W(e^{-1}) \approx -1.278$, where $W$ is the Lambert W-function; the boundaries in (F) are the curves $\beta(E-\mu) = \log z_1^* \approx -1.541$ and $\beta(E-\mu) = \log z_2^* \approx 1.545$, where $z_1^*$, $z_2^*$ are the solutions to $\partial\tilde{s}/\partial\beta = 0$ (Eq.~\ref{eq:dsdbeta}) with $f_0 = 0.99$.
}
\label{fig:selection}
\end{figure}

	     We can also ask how variations in $\beta$ affect the local strength of selection. Variation of $\tilde{s}(E)$ with $\beta$ depends qualitatively on both $E-\mu$ and whether $f_0$ is zero or nonzero. In Fig.~\ref{fig:selection} we show $\log\F(E)$, $\tilde{s}(E)$, and the derivative
     
\beq
\frac{\partial\tilde{s}}{\partial\beta} = \frac{z(1-f_0)}{(1+z)^2(1+f_0z)^2}[(1-f_0z^2)\log z + (1+z)(1+f_0z)].
\label{eq:dsdbeta}
\eeq

\noindent For $f_0 = 0$ (Fig.~\ref{fig:selection}A--C), increasing $\beta$ increases selection strength for $E-\mu \geq 0$.  Here the fitness function drops to zero exponentially, and increasing $\beta$ steepens the exponential drop.  However, for $E-\mu < 0$, the effect of changing $\beta$ depends on the value of $\beta$.  For large $\beta$, increasing $\beta$ actually decreases selection strength; this is because $\beta$ sets the rate at which the Fermi-Dirac function converges to unity, and hence increasing $\beta$ flattens the landscape in that region.  However, for sufficiently small $\beta$, the  threshold region is large enough that increasing $\beta$ still increases selection.  The boundary between positive and negative values of $\partial \tilde{s}/\partial\beta$ are the solutions of the equation $\partial \tilde{s}/\partial\beta = 0$: $\beta (E-\mu) = \log W(e^{-1}) \approx -1.278$, where $W$ is the Lambert W-function (Fig.~\ref{fig:selection}C).

     This situation changes qualitatively in the regime $E-\mu > 0$ when $f_0 \neq 0$ (Fig.~\ref{fig:selection}D-F).  In this case, for sufficiently large $\beta$, increasing $\beta$ weakens selection.  This is different in the case of nonzero $f_0$ because on the high-energy tail, the fitness is converging to a nonzero number $f_0$, and thus selection becomes asymptotically neutral. Hence, when $f_0 \neq 0$, increasing $\beta$ only strengthens selection very close to $E-\mu = 0$.  Using Eq.~\ref{eq:dsdbeta}, the boundaries in Fig.~\ref{fig:selection}F are given by the solutions of $(f_0 z^2 - 1)\log z = (1+z)(1+f_0z)$. This equation can be solved numerically
to obtain two solutions, $z^*_1 < 1$ and $z^*_2 > 1$. The boundaries in Fig.~\ref{fig:selection}F are thus given by the curves $\beta(E-\mu) = \log z^*_1$ for $E-\mu < 0$ and $\beta(E-\mu) = \log z^*_2$ for $E-\mu > 0$.

\subsection{Assessment of model assumptions}

     Two main assumptions inherent in our evolutionary model are monomorphism and steady state.  Here, we assess how violating these assumptions affects inference of evolutionary parameters $\beta$, $\mu$, $\nu$, and $f_0$.  To test this, we generate simulated data sets of binding site sequences evolving under a haploid asexual Wright-Fisher model with the Fermi-Dirac fitness function (Eq.~\ref{eq:FD_fitness}; see Methods for details).

\subsubsection{The effect of polymorphism}

\begin{figure}[ht!]
\centering\includegraphics[width=\textwidth]{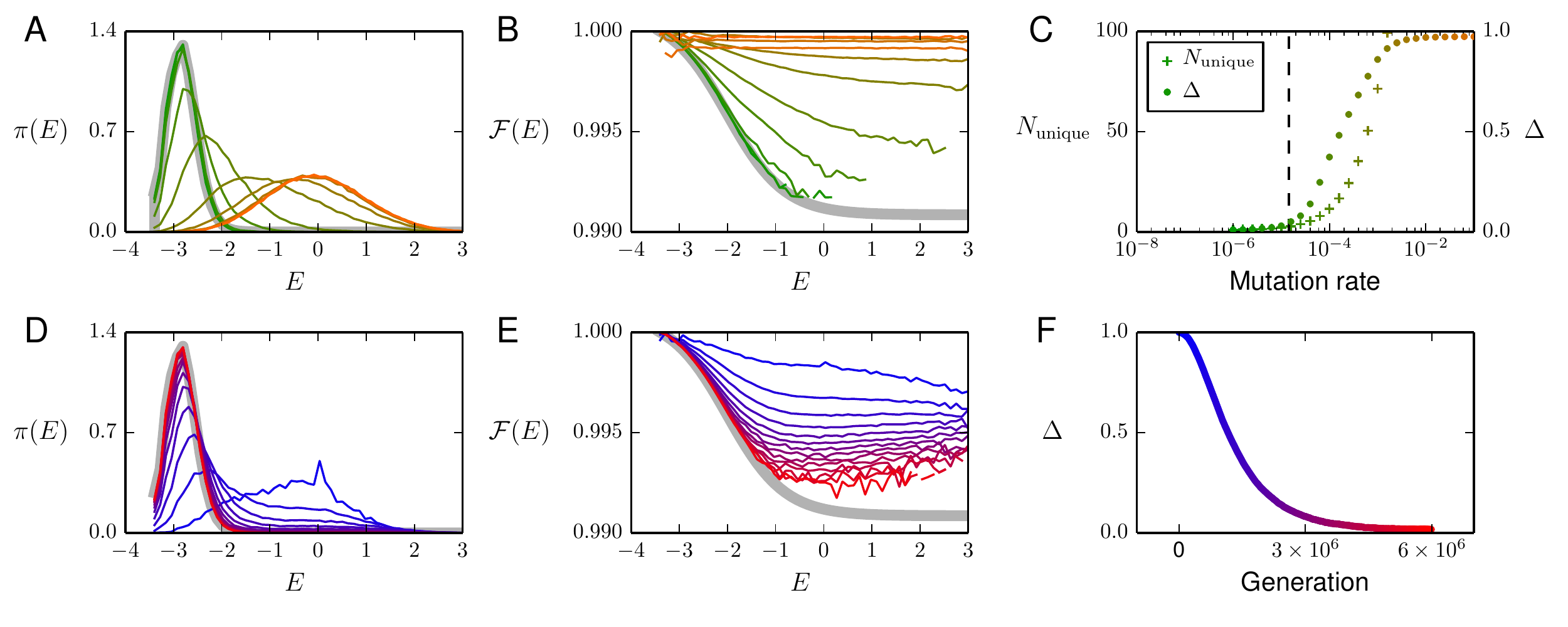}
\caption{The monomorphic limit and steady state of a Wright-Fisher model of population genetics. In (A)--(C) we show results from simulations at various mutation rates, using a fitness function with $f_0 = 0.99$, $\beta = 1.69~(\text{kcal/mol})^{-1}$, and $\mu = -2~\text{kcal/mol}$. Each mutation rate data point is an average over $10^5$ independent runs, as described in Methods. Colors from green to orange correspond to increasing mutation rates.
(A) Observed steady-state distributions $\pi_\text{obs}(E)$ for various mutation rates.  The steady state $\pi(E)$ predicted using Eq.~\ref{eq:steady_state} is shown in grey.
(B) Fitness functions $\F(E)$ predicted using observed distributions $\pi_\text{obs}(E)$ in Eq.~\ref{eq:inference}.  The exact fitness function is shown in gray. Inferred fitness functions are matched to the exact one by using the known population size $N_{e}$, and setting the maximum fitness to 1.0 for each curve.
(C) For each mutation rate, the total variation distance (TVD) $\Delta$ between $\pi_\text{obs}(E)$ and $\pi(E)$, and the average number of unique sequences in the population $N_{\text{unique}}$ (the degree of polymorphism) are shown.  The predicted bound $(N_{e} L \log N_{e})^{-1}$ on mutation rate required for monomorphism is shown as a dashed line.
In (D)--(F) we show simulations in the monomorphic regime which have not reached equilibrium, with the same parameters as in (A)--(C) and $u = 10^{-6}$. Colors from blue to red correspond to the increasing number of generations. In (F), TVD $\Delta$ is calculated in energy space as described in Methods.
}
\label{fig:monomorphic_steady_state_limit}
\end{figure}

     To test the effects of polymorphism on the accuracy of our predictions, we perform a set of simulations for a range of mutation rates $u$. Each simulation in the set follows the Wright-Fisher process to the steady state. We construct the observed distribution $\pi_\text{obs}$ by randomly choosing a single sequence from the final population of each simulation, which may not be monomorphic for larger $u$ (Fig.~\ref{fig:monomorphic_steady_state_limit}A).
From $\pi_\text{obs}$, we infer the fitness landscape as a function of energy 
using Eq.~\ref{eq:inference} 
(Fig.~\ref{fig:monomorphic_steady_state_limit}B).

Additionally, for each $u$ we record the average number of unique sequences present in the population at equilibrium, and
compute the total variation distance (TVD; Eq.~\ref{eq:TVD}) between $\pi_\text{obs}$ and the
monomorphic prediction (Fig.~\ref{fig:monomorphic_steady_state_limit}C).
As expected, at low mutation rates the steady-state distribution and the fitness function match monomorphic predictions well. At higher mutation rates, the TVD starts to increase and
Eq.~\ref{eq:steady_state} overestimates the fitness of low-affinity sites.
The population becomes distinctly polymorphic in this limit.
With very high mutation rates, $\pi_\text{obs}$
approaches the neutral distribution $\pi_0$ since the population is largely composed of newly generated mutants which have not experienced selection.

     A condition for monomorphism in a neutrally evolving population is $u \ll (LN_e \log N_e)^{-1}$~\cite{Champagnat2006}. Indeed, in the monomorphic limit the expected time between new mutations, $(LN_{e}u)^{-1}$, must be longer than the expected time over which fixation occurs, which is ${\cal O} (N_{e})$ generations with probability $1/N_{e}$ for mutants that fix, and $\mathcal{O}(\log N_{e})$ with probability $(N_{e} - 1)/N_{e}$ for mutants that go extinct.  Thus the total expected time before the mutant either fixes or goes extinct is ${\cal O} (\log N_{e})$ generations for $N_{e} \gg 1$~\cite{Kimura1969}. Thus we must have $(LN_{e}u)^{-1} \gg \log N_{e}$ or, equivalently, $u \ll (LN_{e}\log N_{e})^{-1}$.  Using $N_{e}=1000$ and $L=10$ as in our simulations yields $u \ll 1.4 \times 10^{-5}$ in the monomorphic limit, consistent with the results in Fig.~\ref{fig:monomorphic_steady_state_limit}C.
     
\begin{figure}[ht!]
\centering\includegraphics[width=\textwidth]{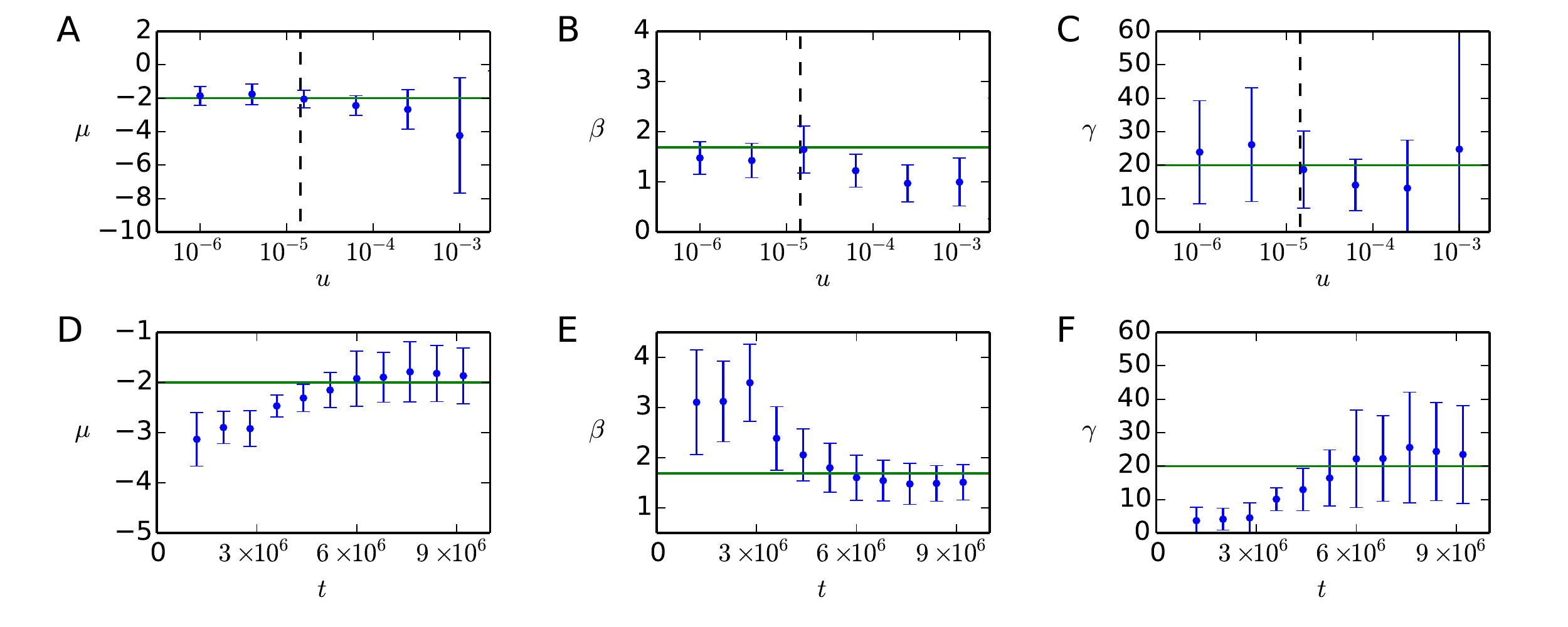}
\caption{Fitted parameters of the Fermi-Dirac function from Wright-Fisher simulations.  In (A)--(C) the fitted values of $\mu$, $\beta$ and $\gamma = \nu(1-f_0)$ are shown as functions of mutation rate $u$.  For each mutation rate, we generate 200 random samples of 500 sequences from the $10^5$ sequences generated in simulations used in Fig.~\ref{fig:monomorphic_steady_state_limit}A--C.  We fit the parameters of the fitness function on each sample separately by maximum likelihood (see Methods). Shown are the averages (points) and standard deviations (error bars) over 200 samples at each mutation rate. The exact values used in the simulation are represented by horizontal green lines. The predicted bound $(N_{e} L \log N_{e})^{-1}$ on mutation rates required for monomorphism is shown as a vertical dashed line.
In (D)--(F) the fitted values of $\mu$, $\beta$, and $\gamma$ are shown as functions of the number of generations $t$, for the equilibration simulations used in Fig.~\ref{fig:monomorphic_steady_state_limit}D--F.  The sampling procedure, the maximum likelihood fit, and the representation of parameter predictions are the same as in (A)--(C).
}
\label{fig:sim_prm_fits}
\end{figure}

We also infer parameters $\beta$, $\mu$ and $\gamma$ with a maximum likelihood fit. As expected, all parameters converge to the exact values in the monomorphic limit (Fig.~\ref{fig:sim_prm_fits}A--C).  When the population is not truly monomorphic, $\mu$ and $\beta$ tend to be underestimated on average, with larger variation in inferred values (larger error bars in Fig.~\ref{fig:sim_prm_fits}A,B).  For $\gamma$, polymorphism has no clear bias on the average inferred value, although it also appears to increase the variation.

\subsubsection{Evolutionary steady state}

     We perform another set of simulations to test the accuracy of our predictions in a population that has not yet reached steady state.  We use the same fitness landscape and population size, but fix $u$ to $10^{-6}$, within the monomorphic limit. At each point in time (measured as the number of generations), we construct $\pi_\text{obs}$ as described in Methods (Fig.~\ref{fig:monomorphic_steady_state_limit}D), and infer the fitness
function (Fig.~\ref{fig:monomorphic_steady_state_limit}E). We also compute the TVD between the observed distribution $\pi_\text{obs}$ and the steady-state prediction (Fig.~\ref{fig:monomorphic_steady_state_limit}F).  With time, $\pi_\text{obs}$ converges to the steady state (Eq.~\ref{eq:steady_state}) and the TVD decays to zero, enabling accurate reconstruction of the fitness function in the threshold region (although it still diverges from the exact function in the high-energy tail, where few sequences are available at steady state).  The equilibration time is expected to be proportional to $u^{-1}$, or $10^6$ generations; indeed, Fig.~\ref{fig:monomorphic_steady_state_limit}F places the equilibration timescale at about $4 \times 10^6$ generations.  As the population equilibrates, accurate inference of the fitness function parameters becomes possible (Fig.~\ref{fig:sim_prm_fits}D-F).  We see that parameters inferred from a population out of steady tend to underestimate $\mu$ and $\gamma$ and overestimate $\beta$.

  

\section{Transcription factor binding sites in yeast}

     
     How well does \emph{S. cerevisiae} satisfy the assumptions of our evolutionary model? \emph{S. cerevisiae} is not a purely haploid organism but goes through both haploid and diploid stages. In \emph{S. paradoxus}, most of the reproduction is haploid and asexual with 1000 generations spent in the haploid stage for each generation in the diploid stage, and heterozygosity is low~\cite{Tsai2008}.  Based on the analysis of yeast genomes, wild yeast populations show extremely limited outcrossing and recombination and are geographically distinct~\cite{Dujon2010}.  Thus, \emph{S. cerevisiae} may be regarded as haploid to a reasonable approximation, with recombination during the diploid stages unlinking the loci. This is consistent with our model, which assumes a haploid population and independent evolution of binding sites.

	Are natural populations of \emph{S. cerevisiae} within the mutation rate limits required for monomorphism?  The mutation rate for \emph{S. cerevisiae} has been estimated to be $0.22 \times 10^{-9}$ mutations per bp per cell division~\cite{Tsai2008}.  Assuming loci of length $L=10$, this sets a bound on the effective population size $N_e$ of $2.7 \times 10^7$, below which the population will be monomorphic. This is roughly equal to the estimated effective population size of \emph{S. cerevisiae} of $ \approx 10^7$ individuals~\cite{Tsai2008}, based on the analysis of neutral regions in the yeast genome. Thus it is plausible that \emph{S. cerevisiae} population sizes are below or near the limit for monomorphism, justifying the use of Eq.~\ref{eq:steady_state}.
Furthermore, in \emph{S. cerevisiae} and \emph{S. paradoxus} the proportion of polymorphic sites in a population has been found to be about 0.001~\cite{Tsai2008, Liti2009, Doniger2008}, generally with no more than two alleles segregating at any one site~\cite{Tsai2008}.  According to this
estimate, we expect about 1\% of binding sites of length 10 bp to be polymorphic, corresponding to an average polymorphism of 1.01 in Fig.~\ref{fig:monomorphic_steady_state_limit}C.
     
    For \textit{S. cerevisiae}, the equilibration time estimate is $u^{-1} \approx 5 \times 10^9$ generations, or about $2 \times 10^6$ years for an estimated 8 generations per day~\cite{Fay2005}. This is several times less than the 5--10 million years of divergence time for the most recent speciation event, with \textit{S. paradoxus}~\cite{Replansky2008}. Thus steady state may plausibly be reached for a fast-reproducing organism like \textit{S. cerevisiae} over evolutionary times scales.


\subsection{Site-specific selection}

Besides the assumptions of monomorphism and steady state, we also require a set of binding sites evolving under universal selection constraints if we are to infer the fitness landscape using Eq.~\ref{eq:inference}. A collection of sites binding to the same TF is an obvious candidate, since these sites all experience the same physical interactions with the TF.  However,
it is possible that selection is site-specific: rather than evolving on the same fitness landscape, different sites for the same TF may be under different
selection pressures depending on which genes they regulate, their 
position on the chromosome, etc. For example, genes under strong selection might require very reliable regulation, so that their upstream binding sites
are selected for tight binding to TFs. In less essential genes,
the requirement of high-affinity binding might be relaxed.
Before directly applying the evolutionary model, we investigate several of these site-specific scenarios to determine if any are supported by the data.
We perform several direct tests of site-specific selection by searching for correlations between site TF-binding energies and other properties of the site
or the gene it regulates.

     We classify fitness effects of genes using knockout lethality, which is available in the Yeast Deletion Database~\cite{Winzeler1999, Giaever2002}. This database classifies genes as either essential or nonessential based on the effects of gene knockout, and provides growth rates for nonessential gene knockouts under a variety of experimental conditions.  We divide binding sites of each TF in our data set into two groups: those regulating essential genes and those regulating nonessential genes.
     
\begin{figure}[ht!]
\centering\includegraphics[width=\textwidth]{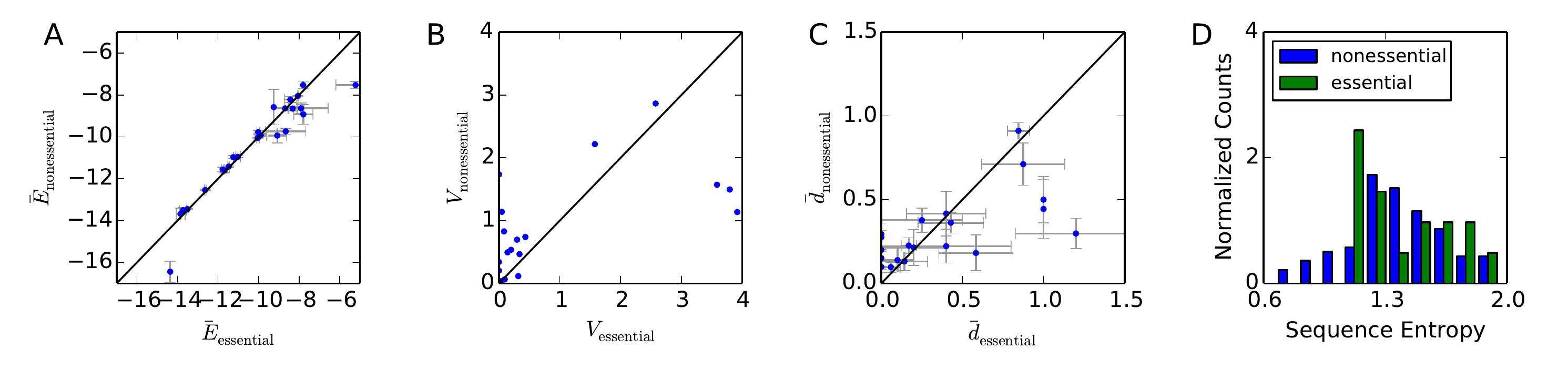}
\caption{Tests of site-specific selection. We divide binding sites for each TF into two groups: those regulating essential and nonessential genes.
(A)~Comparison of mean binding energies of sites regulating essential ($\bar{E}_\text{essential}$) and nonessential genes ($\bar{E}_\text{nonessential}$) for each TF in the data set. Vertical and
horizontal error bars show the standard error of the mean in each group. Points lacking error bars have only one sequence in that group.
(B)~Comparison of variance in binding energies for sites regulating essential ($V_\text{essential}$) and nonessential ($V_\text{nonessential}$) genes.
(C)~Mean Hamming distance between corresponding sites in \emph{S.~cerevisiae} and \emph{S.~paradoxus} for sites regulating essential versus nonessential genes. Vertical and horizontal error bars show the standard error of the mean
in each group. In (A)--(C), 25 TFs were used;
black diagonal lines have slope one.
(D)~Normalized histogram of TF binding site sequence entropies, divided into 16 essential and 109 nonessential TFs, for 125 TFs in Ref.~\cite{Chen2010}.
}
\label{fig:essentiality}
\end{figure}

     In Fig.~\ref{fig:essentiality}A we compare mean binding energies of sites regulating essential genes with those regulating nonessential genes for each TF.  Using a null model as described in Methods, we find no significant difference (at $p = 0.05$ level) between the two groups of sites for any TF except RPN4, for which $p=0.03$ and the difference in mean energies is 0.24 kcal/mol, and PDR3, for which $p=0.002$ and the difference in mean energies is 2.3 kcal/mol.
The mean $p$-value of the null model over all TFs is 0.38. In Fig.~\ref{fig:essentiality}B we compare the variance of the energy of the sites
regulating essential and nonessential genes; sites regulating essential genes may be selected for more specific values of binding energy if precise regulation is required.  We find no overall trend: for some TFs sites regulating essential genes have more energy variation than those regulating nonessential genes, but for other TFs the situation is reversed.
     
     For the sites regulating nonessential genes, we also correlate the site binding energy with the growth rate of a strain in which the regulated gene was knocked out (Table S1, column B). The Spearman rank correlation between each site's binding energy and the regulated gene's effect on growth rate produces a mean $p$-value of 0.51. We find no significant correlation for any TF at $p=0.05$ level except MSN2, with $p=0.046$.
 
     It is possible that regulation of highly-expressed genes may be more tightly controlled.  Indeed, gene expression level is weakly, though significantly, correlated with gene essentiality~\cite{Holstege1998}. We compare the binding energy of sites to the overall expression level of their regulated genes measured in mid-logphase yeast cells cultured in YPD~\cite{Holstege1998} (Table S1, column C), and again find no correlation using the Spearman rank correlation except for DAL80 ($p=0.034$), with mean $p$-value of 0.54.
     
     Another measure of the selection pressures on genes is their rate of evolution as measured by $K_A/K_S$, the ratio of nonsynonymous to synonymous mutations in a given gene between species.  According to the neutral theory of evolution, genes which evolve slowly must be under higher selective pressure, and therefore the sites regulating them might likewise experience stronger selective pressures.  As described in Methods, we measure the $K_A/K_S$ ratio between \emph{S.~cerevisiae} and \emph{S.~paradoxus} protein coding sequences, and compare it to the binding energy of the sites regulating those genes (Table S1, column D).  We find very weak Spearman rank correlations for ATF2, RPN4, GAT1 and CAD1 all roughly with $p=0.02$. We find no other significant correlation at the $p=0.05$ level, with a mean $p$-value of 0.42.

     Similarly, one might expect sites regulating essential genes to be more conserved. However, we find that the average Hamming distance between corresponding binding sites in \emph{S.~cerevisiae} and \emph{S.~paradoxus}~\cite{Chen2010} is no different for sites regulating essential genes than for those regulating nonessential genes, as shown in Fig.~\ref{fig:essentiality}C. Using the null model described in Methods, most TFs are above $p=0.05$ with the exceptions of YAP7 ($p=0.04$) and PDR3 ($p=0.003$), with an average $p$-value of 0.27. 
     
     We can also consider how the essentiality of the TFs themselves affects the sequences of their binding sites; for example, essential TFs may constrain their binding sites to a more conserved sequence motif. We divide 125 TFs from Ref.~\cite{Chen2010} which had 10 or more sequences and for which essentiality information was available into 16 essential and 109 nonessential TFs using the Yeast Deletion Database~\cite{Winzeler1999, Giaever2002}, and calculate the sequence entropy of binding sites for each TF. The distribution of sequence entropies in Fig.~\ref{fig:essentiality}D shows no significant difference between essential and nonessential TFs ($p=0.9$ for the null model).

Finally, it is possible that sites experience different selection pressures depending on their distance to the transcription start site (TSS).
Again, we find no significant correlations between binding energy and distance to the TSS: Spearman rank correlation yields mean $p$-value of 0.59 and all $p$-values above 0.05 (Table S1, column E).
Overall, our findings are in broad agreement with a previous report~\cite{Mustonen2008}, which suggested that site-specific selection can be ruled out because of the significant variation in binding affinity between orthologous sites of different species, which is consistent with the variance predicted by a model including only drift and site-independent selection.


\subsection{Inference of biophysical fitness landscapes}

     The above analysis indicates that the evolution of binding site energies does not depend significantly on site-specific effects, suggesting that more universal principles govern the observed distribution of sites binding a given TF.  Thus, we can fit a single fitness function to a collection of TF-bound sites via Eqs.~\ref{eq:steady_state} and~\ref{eq:inference}. Of the 25 TFs considered in the previous section, here we focus on 12 TFs with $>12$ unique binding site sequences.

\begin{figure}[ht!]
\centering\includegraphics[width=\textwidth]{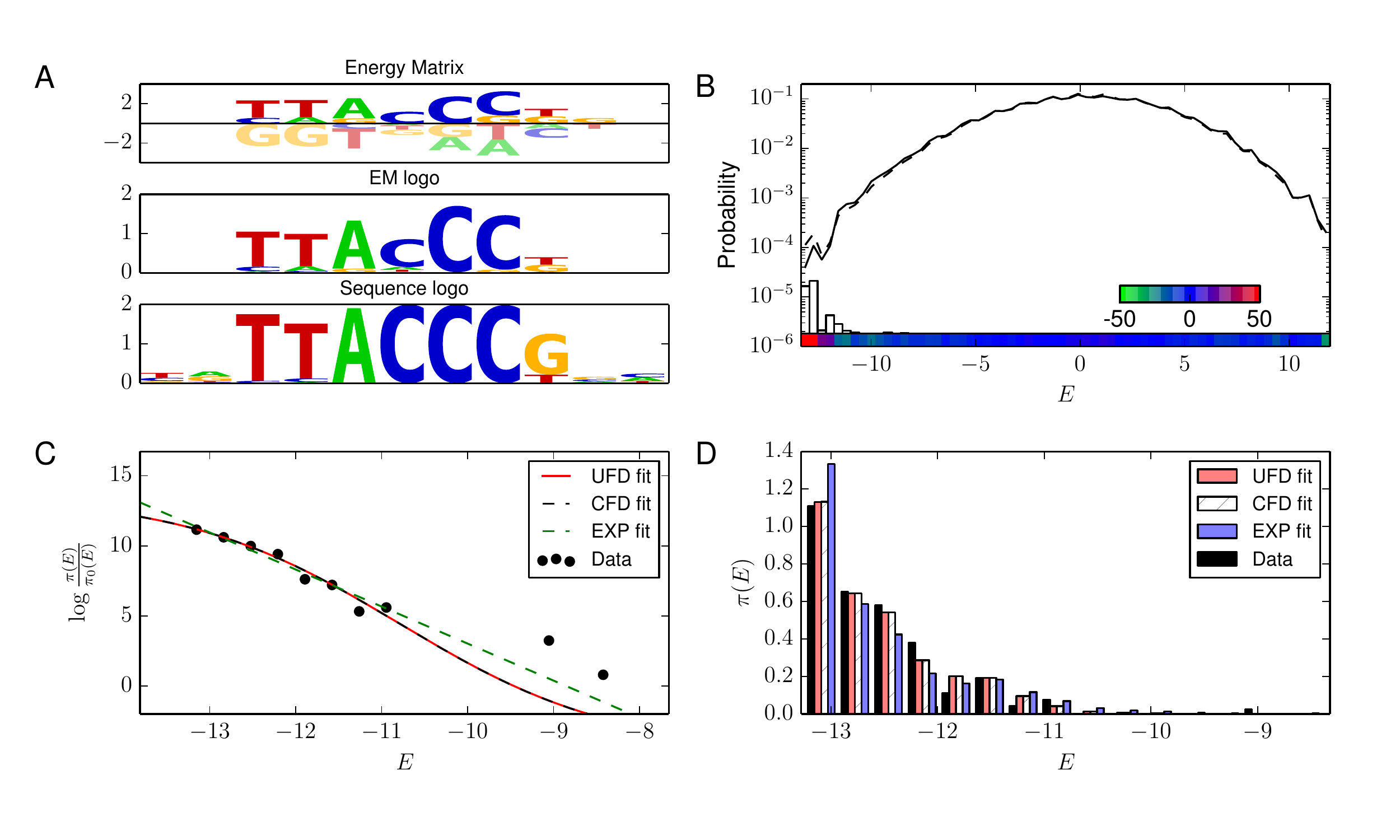}
\caption{Parametric inference of REB1 fitness landscape.
(A)~From top to bottom: REB1 EM~\cite{Fordyce2010}, the sequence logo
obtained from the EM by assuming a Boltzmann distribution at room temperature
at each position in the binding site ($\pi^i (\s_i) = \pi_0^i (\s_i) e^{-\beta \epsilon_i^{\s_i}}/Z_i$), and the sequence logo based on the
alignment of REB1 genomic sites.
(B)~Histogram of energies of intergenic sites calculated using the REB1 EM (dashed line).  The neutral distribution of sequence energies expected from the mono- and dinucleotide background model (solid line; see Methods for details).
The histogram shows the distribution of functional sites~\cite{Chen2010}.
The color bar on the bottom indicates the percent deviation between the two distributions (red is excess, green is depletion relative to the background model).
(C)~Fitness function inference. Dots represent data points (as in Fig.~\ref{fig:raw_fitness}); also shown are the unconstrained fit to the Fermi-Dirac function of Eq.~\ref{eq:FD_fitness} (``UFD''; solid red line), constrained fit to the Eq.~\ref{eq:FD_fitness} with $f_0 = 0.99$ (``CFD''; dashed black line), and fit to an exponential fitness function (``EXP''; dashed green line).
(D)~Histogram of binding site energies and its prediction based on the three fits in (C) (Eq.~\ref{eq:energy_proj}).}
\label{fig:REB1_fit}
\end{figure}

     First we derive the neutral distribution $\pi_0(E)$ of site energies based on mono- and dinucleotide frequencies obtained from intergenic regions of the \emph{S.~cerevisiae} genome, as described in Methods.  It has been suggested that $L$-mers not functioning as regulatory sites (e.g., located outside promoters) may be under evolutionary pressure not to bind TFs~\cite{Hahn2003}; however, consistent with previous reports~\cite{Djordjevic2003,Mustonen2008}, we find that sequences sampled from the intergenic regions of the genome are close to the neutral distribution expected from mono- and dinucleotide frequencies, except for the expected enrichment at low energies due to functional binding sites.  This distribution is shown in Fig.~\ref{fig:REB1_fit}A for REB1 and in Table S2,
column B for all other TFs.

\begin{figure}[ht!]
\centering\includegraphics[width=\textwidth]{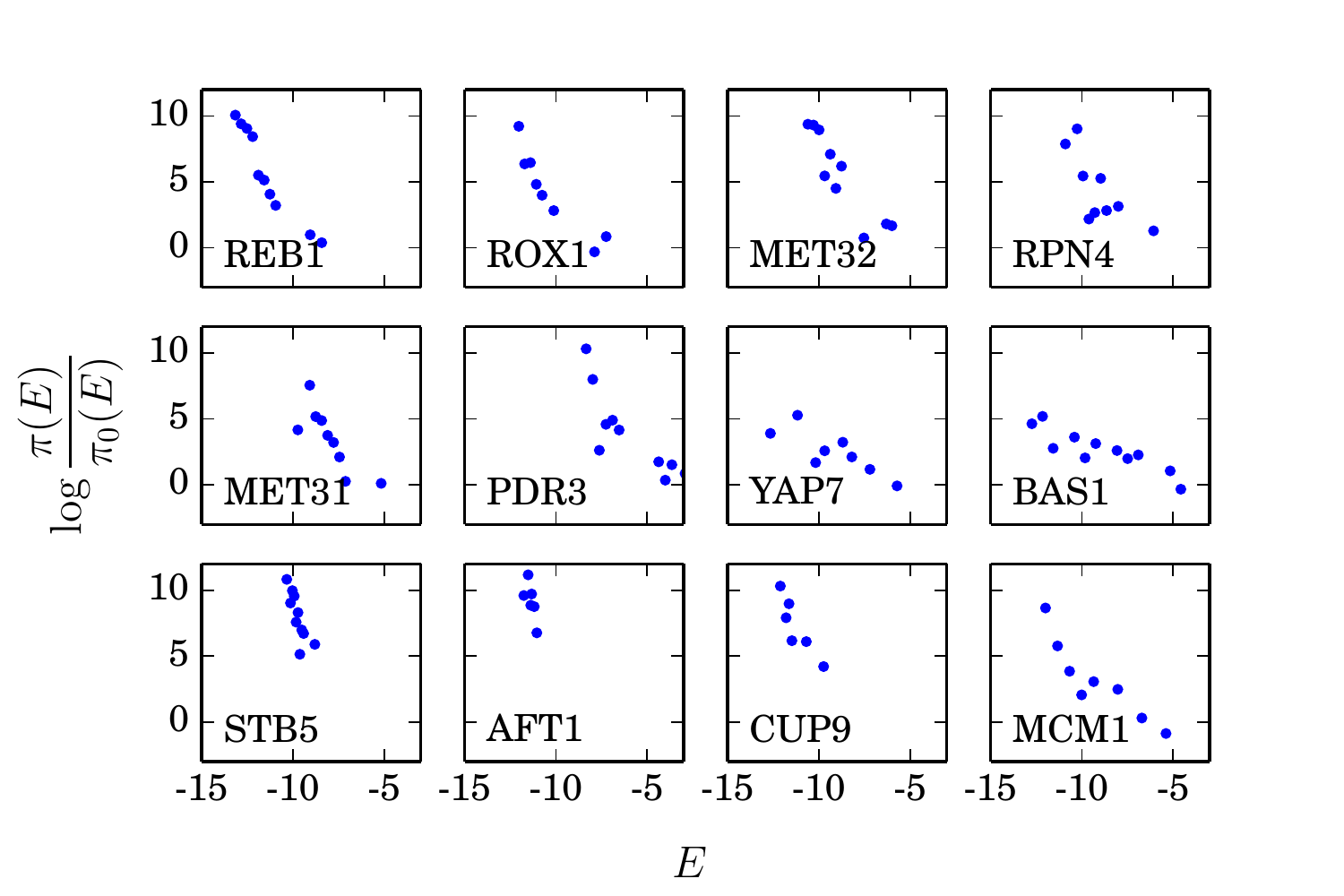}
\caption{Qualitative behavior of fitness landscapes. Shown are plots of $\log(\pi(E)/\pi_0(E))$ for 12 TFs, which, according to Eq.~\ref{eq:inference}, equals the logarithm of fitness up to an overall scale and shift. For each TF, sequences are grouped into 15 equal-size energy bins between the minimum and maximum energies allowed by the EM.}
\label{fig:raw_fitness}
\end{figure}

     Assuming the observed set of binding site energies for a TF adequately samples the distribution $\pi(E)$, we can use our estimate of the neutral distribution $\pi_0(E)$ in Eq.~\ref{eq:inference} to reconstruct the fitness landscapes as a function of TF binding energy up to an overall scale and shift
(Fig.~\ref{fig:raw_fitness}). Although the fitness functions may be noisy
due to imperfect sampling of $\pi(E)$, they nevertheless provide important qualitative insights. In particular, in all landscapes fitness decreases monotonically as binding energy increases, indicating that 
stronger-binding sites are more fit. Moreover, we observe no fitness penalty for
binding too strongly, at least within the range of energies spanned by $\pi(E)$.

\subsubsection{Fermi-Dirac landscapes and model selection} 

     For each TF we perform a maximum-likelihood fit of the binding site data to the distribution in Eq.~\ref{eq:steady_state} with the Fermi-Dirac landscape of Eq.~\ref{eq:FD_fitness} (Fig.~\ref{fig:REB1_fit}, Table S2; see Methods for details). The model of Eq.~\ref{eq:FD_fitness} has four fitting parameters: $\beta$, $\mu$, $\nu$, and $f_0$. However, as shown in Sec.~\ref{sec:biophysical_model},
in the $1-f_0 \ll 1$ limit the fitness function depends on $\gamma = \nu (1 - f_0)$ rather than $f_0$ and $\nu$ separately. Thus we also carry out
constrained ``non-lethal'' Fermi-Dirac fits in which $f_0$ is fixed at 0.99.
Although the inverse temperature $\beta$ and the chemical potential $\mu$ have unambiguous physical meanings in the binding probability of Eq.~\ref{eq:prob_bound}, we will interpret the fits more broadly to define a class of fitness landscapes with ``effective'' $\beta$ and $\mu$, which may not be equal to their physical counterparts.
The input to each fit is a collection of genomic TF binding sites $\{\s\}$~\cite{Chen2010} and the energy matrix from high-throughput \textit{in vitro} TF-DNA binding assays~\cite{Fordyce2010}, which allows us to assign
a binding energy $E(\s)$ to each site.

     A summary of maximum-likelihood parameter values for all TFs is shown in Tables~\ref{table:fits} and S2, column D. The variation of log-likelihood with fitting parameters is shown in Table S2, columns G and H.  Six of the TFs
(REB1, ROX1, MET32, PDR3, CUP9, and MCM1) are in the $1-f_0 \ll 1$ regime
where only $\gamma$ can be inferred unambiguously. Indeed, non-lethal Fermi-Dirac fits with $f_0 = 0.99$ yield
very similar values of log-likelihood and $\gamma$ (Table S2, column D).
In all of these cases, $\gamma$ is considerably greater than $1$, implying that
selection is strong compared to drift, and the effective population size is
large (the $s \ll 1$, $N_e s \gg 1$ regime in population genetics).

     Five TFs (RPN4, MET31, YAP7, BAS1, and AFT1) have very small values of $f_0$ (Table~\ref{table:fits}), indicating that on average, removing their binding sites is strongly deleterious to the cell. In these cases, the degeneracy is broken and the global maximum occurs in the vicinity of $f_0 = 0$ (Table S2, column H, insets). Since $1 - f_0 \approx 1$, $\nu \approx \gamma$, a small value in four out of five cases (Table~\ref{table:fits}). Given the strength of selection, small effective population sizes (which indicate that genetic drift is strong) are necessary to reproduce the observed variation in binding site sequences.  Finally, sites for STB5 have an intermediate value of $f_0 = 0.167$, which means they are under strong selection but are not necessarily essential.

     Since the constrained Fermi-Dirac fits have one less adjustable parameter, it is more consistent to do model selection on the basis of the Akaike information criterion (adjusted for finite-size samples)~\cite{Burnham2002} rather than log-likelihoods:
     
\beq
\text{AIC} = 2(k - \log \mathcal{L}) + \frac{2k(k+1)}{n - k - 1},
\label{AIC}
\eeq

\noindent where $k$ is the number of fitting parameters, $\mathcal{L}$ is the likelihood, and $n$ is the number of data points. For each model we can calculate the AIC, which accounts for both the benefits of higher log-likelihood and the costs of additional parameters.

     Table~\ref{table:aic} shows the AIC differences between the unconstrained Fermi-Dirac fits (UFD, $k=4$) and the constrained Fermi-Dirac fits with $f_0 = 0.99$ (CFD, $k=3$) for each TF.  Positive AIC differences indicate that UFD is more favorable.  We also calculate the Akaike weights $w \propto e^{-\text{AIC}/2}$, which give the relative likelihood that a given model is the best~\cite{Burnham2002}.

     For the six TFs in the $1-f_0 \ll 1$ regime, the constrained Fermi-Dirac fits perform somewhat but not drastically better than the unconstrained Fermi-Dirac fits.  Indeed, the Akaike weights for the constrained Fermi-Dirac fits exceed the full fits for these TFs consistently by about a factor of $e \approx 2.7$, since their raw likelihoods are essentially equivalent and they only differ in the number of fitted parameters $k$.   Out of the five TFs for which $f_0 \approx 0$, YAP7, BAS1, and AFT1 fit slightly better to the constrained Fermi-Dirac, suggesting that their fitted values of $f_0$ are not significant.  For RPN4 and MET31, the AIC analysis shows preference for the fits with low $f_0$.  This preference is especially strong for RPN4 (Table~\ref{table:aic}).  Both RPN4 and MET31 are listed as nonessential in the Yeast Deletion Database~\cite{Winzeler1999, Giaever2002}, suggesting an inconsistency in our analysis.

	The fits to the Fermi-Dirac fitness landscapes also provide estimates of the effective inverse temperature $\beta$ and the effective chemical potential $\mu$ (Table~\ref{table:fits}). The inferred values of $\beta$ can be compared to the physical value at room temperature, $\beta_{ph} = 1.69~ (\text{kcal/mol})^{-1}$.  Nine of the TFs (REB1, ROX1, MET32, PDR3, YAP7, BAS1, STB5, CUP9, MCM1) have $\beta$'s lower than the physical value, while in the other three (RPN4, MET31, AFT1) $\beta > \beta_{ph}$.
In most TFs the fitted inverse temperature $\beta$ is far from its
physical counterpart, although in several cases the likelihood function
is fairly flat in the vicinity of the peak, indicating that a wider range of $\beta$ values is admissible (Table~S2, column G).

	The inferred value of $\mu$ relative to the distribution of energies $E$ of the binding sites tells us which qualitative regime of the Fermi-Dirac fitness landscape the sites lie in.  For five TFs (ROX1, MET32, PDR3, CUP9, MCM1), $E-\mu > 0$, and the sites reside on the exponential tail. Interestingly, $1 - f_0 \ll 1$ for all of these TFs.  For another group of five TFs (REB1, RPN4, MET31, YAP7, AFT1), $E-\mu \approx 0$, so that the sites lie on the sharp threshold between fully bound and fully unbound states.  In this regime, changing the energy of the site through mutations may lead to a large change in its occupancy by the TF.  Finally, for two TFs (BAS1, STB5), $E-\mu < 0$, and the sites lie on the 
high-fitness plateau.  Note that in most of the $E-\mu > 0$ and $E-\mu < 0$ cases, values of $\mu$ within a large range fit the data equally well, as long as the binding energies of all sites are well to the right or to the left of the chemical potential (Table~S2, column G).


What does $\beta \ne \beta_{ph}$ say about the nature and strength of selection?  We address this question using the local selection coefficient, 
$\tilde{s}(E) = |d\log\F/dE|$ (Eq.~\ref{eq:sE}). The magnitude of the selection coefficient depends qualitatively on both $E-\mu$ and whether $f_0$ is zero or nonzero (Fig.~\ref{fig:selection}). For five of the TFs (ROX1, MET32, PDR3, CUP9, MCM1), $f_0 \neq 0$,
$\beta < \beta_{ph}$, and $E-\mu > 0$. Thus these TFs are in a regime
where decreasing $\beta$ strengthens selection (Fig.~\ref{fig:selection}F).
In other words, selection is stronger for these binding sites than expected from purely biophysical considerations. For RPN4, MET31, and AFT1, $f_0 \approx 0$, $\beta > \beta_{ph}$, and $E \approx \mu$. Hence $\partial\tilde{s}/\partial\beta > 0$, and selection is again stronger than expected.  BAS1 and STB5 exhibit $\beta < \beta_{ph}$ and lie on the high fitness plateau ($E-\mu < 0$), and thus selection is also stronger than expected.  In contrast, YAP7 and REB1 exhibit $\beta < \beta_{ph}$ and lie on the threshold $E - \mu \approx 0$, and hence selection is weaker than expected
in these two cases.

\subsubsection{Exponential fitness landscape}
     
     Next, we consider a purely exponential fitness landscape of the form $\F(E) = e^{\alpha E}$.  The reasons for including this case are threefold.  First, exponential fitness emerges in the limit $E-\mu \gg 0$ of the Fermi-Dirac landscape, the regime into which many of the TF binding sites fall. Second, the fitness landscapes in Fig.~\ref{fig:raw_fitness} appear close to linear on the logarithmic scale, implying that to a good approximation fitness depends exponentially on energy.  Third, the model has just one fitting parameter.
     
The steady-state distribution $\pi(\s)$ with exponential fitness is given by

\beq
\begin{split}
\pi(\s) & = \frac{1}{Z} \pi_0(\s) e^{\nu\alpha E(\s)} \\
& = \prod_{i = 1}^L \frac{\pi_0^i(\s_i)}{Z_i} e^{\nu\alpha \epsilon^{\s_i}_i},
\end{split}
\eeq

\noindent where $E(\s)$ is given by Eq.~\ref{eq:seq_energy}, $\pi_0(\s)$ 
is the neutral probability of sequence $\s$, $\pi_0^i(\s_i)$ is
the background probability of nucleotide $\s_i$ at position $i$,
and $Z_i$ is a single-site partition function:
$\pi_0(\s)/Z = \prod_{i=1}^L \pi_0^i(\s_i)/Z_i$.
Here we assumed that the background probability of a sequence is a product
of probabilities of its constituent nucleotides.
In this case, sites decouple and the distribution of sites
$\pi(\s)$ completely factorizes.  
The assumption of factorization underlies the common practice of inferring EMs from log-odds scores of observed genomic binding sites~\cite{Stormo1998}. The log-odds score of a nucleotide $\s_i$ is defined as

\beq
{\cal S} (\s_i) = \log \frac{p_i^{\s_i}}{\pi_0^i(\s_i)} = -\beta \epsilon_i^{\s_i} - \log{Z_i},
\label{eq:log_odds}
\eeq

\noindent where $p_i^{\s_i}$ is the probability of seeing base $\s_i \in \{\mathsf{A,C,G,T}\}$ at position $i$ within the set of known sites, $\beta$ is an effective inverse temperature, and $Z_i$ is the normalization constant.
Eq.~\ref{eq:log_odds} shows that the log-odds score, which is computed using observed nucleotide probabilities, is equivalent to $\epsilon_i^{\s_i}$ (up to an overall scale and shift) under the assumption of site-independence.

     We can quantitatively compare the exponential fitness landscape with the unconstrained and constrained Fermi-Dirac landscapes using the Akaike information criterion, Eq.~\ref{AIC}. The AIC analysis shows that the exponential landscape is significantly poorer than the Fermi-Dirac landscape in all cases except STB5 (Table~\ref{table:aic}). This observation provides
statistical support for the fitness landscapes of Fermi-Dirac type,
and for the non-lethality of deleting most TFs (the exponential fitness decays to zero rather than a nonzero $f_0$ found in most of our Fermi-Dirac fits).


\section{Discussion}

In this work, we have considered how fitness of a single-cell eukaryote \textit{S. cerevisiae} 
is affected by interactions between TFs and their cognate genomic sites.
Changing the energy of a site, or creating new sites in gene
promoters may change how genes are activated and repressed, which
in turn alters the cell's chances of survival.
Under the assumptions of a haploid monomorphic population in which evolution
of binding sites has reached steady state, the fitness landscape as a function 
of TF binding energy can be inferred from the distribution of TF binding sites
observed in the genome, using a biophysical model which assigns
binding energies to sites.  We use a simple EM model of TF-DNA energetics
in which the energy of each position in the site is independent of all the other
positions. The EM parameters are inferred from a high-throughput
data set in which TF-DNA interactions were studied \textit{in vitro} using
a microfluidics device~\cite{Fordyce2010}. We consider two types of fitness
functions: Fermi-Dirac, which appears naturally from considering TF binding
as a two-state process (Eq.~\ref{eq:prob_bound}), and exponential, which
is motivated by the observation that for many TFs, fitness appears to fall off linearly with energy in log-space.

A single fitness landscape for all genomic binding sites of a given TF can only
exist in the absence of site-specific selection. Indeed, it is
possible that TF sites experience different selection pressures depending on
the genes they regulate: for example, sites in promoters of essential genes
may be penalized more for deviating from the consensus sequence. In this
case, the fitness function is an average over all sites which evolve under different selection constraints: as an extreme example, consider the case where each site $i$ has a Fermi-Dirac fitness function (Eq.~\ref{eq:FD_fitness}) with different parameters $\mu_i$, $\beta_i$, and $f_{0,i}$. The resulting observed distribution of energies would then be the average of the distributions predicted by Eq.~\ref{eq:energy_proj}:

\begin{equation}
\pi(E) = \frac{1}{Z} \pi_0(E) \langle \F(E; \mu_i, \beta_i, f_{0i})^\nu  \rangle_i \equiv \frac{1}{Z} \pi_0(E) \F (E; \bar{\mu}, \bar{\beta}, \bar{f_0})^{\bar{\nu}},
\end{equation}
which defines the ``average'' fitness function with effective parameters $\bar{\mu}$, $\bar{\beta}$, $\bar{f_0}$, $\bar{\nu}$. Thus the fit can be carried out even in the presence of site-dependent selection, but the fitted parameters correspond to fitness functions of individual sites only in an average sense.

In order to gauge the importance of site-specific selection in TF binding site
evolution, we have performed several statistical tests aimed at discovering
correlations between binding site energies and biological properties of the sites and the genes they regulate. These tests considered gene essentiality,
growth rates of strains with nonessential genes knocked out, gene expression
levels, $K_A/K_S$ ratios based on alignments with \textit{S.paradoxus}, and
the distance of the site to the TSS.
The results of these tests indicate that for a given TF,
the evolution of regulatory sites is largely independent of the properties of regulated genes and the specific biological functions of the sites.

Previously, low correlations have been observed between essentiality and conservation of protein and coding sequences~\cite{Jordan2002, Pal2003, Zhang2005, Choi2007, Krylov2003, Wang2009, Fang2005}, which has fueled considerable speculation as it contradicts the prediction of the neutral theory of evolution that higher selection pressures lead to lower evolutionary rates. It has also been found that the growth rate of strains with nonessential genes knocked out is significantly (though weakly) correlated with conservation of those genes~\cite{Hirsh2001}.  It has therefore been suggested that selection pressures are so strong that only the most nonessential genes experience significant genetic drift~\cite{Jordan2002}. Previous studies have also found that gene expression levels are a more reliable (though still very weak) predictor of selection pressures than essentiality~\cite{Krylov2003}, but we do not find this to be the case for TF binding sites, nor do we observe a significant correlation between gene expression levels and TF binding energies.


	Available data does not rule out the possibility of time-dependent selection in combination with forms of site-dependent selection we have not accounted for. In this scenario, the variation in site binding affinity is not due to genetic drift, but to variable selection pressures across sites and over time,
such that the sites are strongly tuned to particular binding energies which change from locus to locus. Indeed, there is evidence that there is frequent gain and loss of TF binding sites and that the gene regulatory network is highly dynamic~\cite{Doniger2007, Raijman2008, Tirosh2008, Tuch2008, Jovelin2009, Wuchty2003, vanDijk2012}.
However, it is possible that rapid turnover of binding sites in eukaryotes may be due to evolution acting on whole promoters rather than individual binding sites. Many promoters contain multiple binding sites for a single TF, and it may be that while individual binding sites are lost and gained frequently, the overall binding affinity of a promoter to a TF may be held constant~\cite{He2012, He2011, Habib2012}.
Our evolutionary model can account for this scenario using a promoter-level fitness function, which we will study in future work.


Out of 12 TFs with sufficient binding site data, five have $f_0 \approx 0$,
indicating a large fitness penalty for deleting such sites. This conclusion is strongly supported by the AIC differences between unconstrained and non-lethal Fermi-Dirac fits for only one TF, RPN4 (Table~\ref{table:aic}). RPN4 is classified as nonessential in
the Yeast Deletion Database. It may be that this misclassification is due to
a mismatch between genomic sites, in which the core $\mathsf{GCCACC}$ motif is preceded by $\mathsf{TTT}$, and the EM in which the binding energies upstream of the core motif
are non-specific. We also classify REB1 and MCM1 binding sites as nonessential,
although knocking out these TFs is lethal in yeast. This discrepancy may be due
to a minority of essential sites being averaged with the majority of nonessential sites to produce a single fitness function, as described above.
In addition, although a penalty for deleting any single site may be small,
the cumulative penalty for deleting all sites (or, equivalently, deleting
the TF) may be lethal. Overall, on the basis of AIC we classify 8 out of 12 TFs correctly (Table~\ref{table:aic}).

We find that in 10 out of 12 cases, fitting an exponential fitness function
is less supported by the data than fitting a Fermi-Dirac function (Table~\ref{table:aic}). This is interesting since constructing a position-specific weight matrix by aligning genomic sites is a common practice which
implicitly assumes factorization of exponential fitness and independence of each position
in the binding site. Our results indicate epistasis among positions and
show the limitations of this approximation.

Finally, we find that depending on the TF the distribution of TF binding energies may fall on the exponential tail, across the threshold region, or on the saturated plateau where the sites are always occupied (Table~\ref{table:fits}). In the first two categories, variation of TF concentration in the cell will lead to graded responses, which may be necessary to achieve precise and coordinated gene regulation. In the third regime, TF binding is robust but not dynamic. We also find that the fitted inverse temperature $\beta$ is typically not close to the value based on room
temperature (Table~\ref{table:fits}). This observation suggests selection
pressures in addition to those dictated by the energetics of TF binding to
its cognate sites.


\section{Methods}
\footnotesize


\subsection{The steady state distribution}

     In the limit $u \ll (LN_{e}\log N_{e})^{-1}$, where $u$ is the mutation rate per nucleotide, $L$ is the number of nucleotides in a locus, and $N_{e}$ is the effective population size, mutations are sufficiently rare that each new mutation either fixes or goes extinct before the next one arrives~\cite{Champagnat2006}.  Thus populations evolve by sequential substitutions of new mutations at a locus, which consist of a single new mutant arising and then fixing.  The rate at which a given substitution occurs is thus given by the rate of producing a single mutant times the probability that the mutation fixes~\cite{Kimura1983}:

\beq
W(\s'|\s) \approx N_{e} u(\s'|\s) \cdot \phi(\s' |\s),
\label{eq:rates}
\eeq

\noindent where $N_{e}$ is an effective population size, $u(\s'|\s)$ is the mutation rate from $\s$ to $\s'$, and $\phi(\s' |\s)$ is the probability that a single $\s'$ mutation fixes in a population of wild-type $\s$.  We will assume that $u$ is nonzero only for sequences $\s$ and $\s'$ differing by a single nucleotide.

     Given an ensemble of populations evolving with these rates, we can define $\pi(\s, t)$ to be the probability that a population has sequence $\s$ at time $t$.  This probability evolves over time via the master equation
     
\beq
\frac{d}{dt} \pi(\s', t) = \sum_{\s \in \mathcal{S}} [W(\s'|\s) ~\pi(\s, t) - W(\s| \s') ~ \pi(\s', t)], 
\label{eq:evol_equation}
\eeq

\noindent where $\mathcal{S}$ is the set of all possible sequences at the locus of interest.  This Markov process is finite and irreducible, since there is a nonzero probability of reaching any sequence from any other sequence in finite time.  Hence it has a unique steady-state distribution $\pi(\s)$ satisfying~\cite{Allen2011}

\beq
\sum_{\s \in \mathcal{S}} [W(\s'|\s) ~\pi(\s) - W(\s| \s') ~ \pi(\s')] = 0.
\label{eq:ss_condition}
\eeq


     For population models obeying time reversibility, we can show that the steady-state distribution $\pi(\s)$ must have the form in Eq.~\ref{eq:steady_state}.  We assume the fixation probability $\phi$ depends only on the ratio of mutant to wild-type fitnesses: $\phi(\s'|\s) = \phi(\F(\s')/\F(\s))$.  This occurs in most standard population models and is expected whenever only relative fitness matters (e.g., when the total population size is constant).  If the population dynamics are time reversible, the substitution rates and steady state must obey the detailed balance relation $W(\s'|\s) \pi(\s) = W(\s|\s') \pi(\s')$.  Assuming the neutral dynamics also obey detailed balance, $u(\s'|\s) \pi_0(\s) = u(\s|\s') \pi_0(\s')$, we can show that

\beq
\frac{\pi(\s')}{\pi(\s)} = \frac{u(\s'|\s)}{u(\s|\s')} \frac{\phi\left(\frac{\F(\s')}{\F(\s)}\right)}{\phi\left(\frac{\F(\s)}{\F(\s')}\right)} = \frac{\pi_0(\s')}{\pi_0(\s)} \psi\left( \frac{\F(\s')}{\F(\s)} \right),
\label{eq:psi}
\eeq

\noindent where $\psi(r) = \phi(r)/\phi(1/r)$. Eq.~\ref{eq:psi} implies that $\psi(r) \psi(r') = \psi(r r')$, leading to $\psi(r) = r^\nu$ for some exponent $\nu$.  It can be shown that $\nu$ must be proportional to the effective population size; for the Wright-Fisher model, $\nu = 2(N_{e}-1)$.  Now Eq.~\ref{eq:steady_state} follows from

\beq
\frac{\pi(\s')}{\pi(\s)} = \frac{\pi_0(\s')}{\pi_0(\s)} \left( \frac{\F(\s')}{\F(\s)} \right)^\nu.
\eeq

     This form of the steady state assumes only time reversibility and dependence on fitness ratios; otherwise, any form of the fixation probability must satisfy it.  While many population models do not obey time reversibility exactly, it can be shown that even these irreversible models satisfy Eq.~\ref{eq:steady_state} to a very good approximation~\cite{Manhart2012}.
     

\subsection{Maximum-likelihood fits of fitness function parameters}

     For a given TF, let $S = \{ \s \}$ be the set of binding site sequences, and $\theta = (\beta, \mu, f_0, \nu)$ the parameters of the fitness function (Eq.~\ref{eq:FD_fitness}).  The log-likelihood is
given by

\beq
\log \mathcal{L}(S | \theta) = \sum_{\s \in S} \log \pi(\s |\theta) = \sum_{\s \in S} \log\left( \frac{1}{Z(\theta)} \pi_0(\s) (\F(\s|\theta))^{\nu}\right),
\eeq

\noindent where $\F$ is the fitness function, and
$Z(\theta) = \sum_{\s} \pi_0(\s) (\F(\s|\theta))^{\nu}$ is the normalization.

     Because the log-likelihood function has degenerate or nearly-degenerate regions in the parameter space of $\theta$, we carry out its maximization in two steps.  We first obtain a global map of the likelihood by calculating the function over a mesh of points in parameter space, over the domain $\beta \in (0.1, 10)$, $\mu \in (-20, 0)$, $\nu \in (10^{-3}, 10^5)$, and $f_0 \in (4.5 \times 10^{-5}, 1 - 4.5 \times 10^{-5})$.  We then maximize the likelihood using conjugate-gradient ascent which starts from the approximate global maximum on the mesh.


\subsection{Binding site and EM data}

     We obtain curated binding site locations for 125 TFs from Ref.~\cite{Chen2010}, which provides a posterior probability that each site
is functional based on cross-species analysis.  We only consider sites with a posterior probability above 0.9.
Fro this analysis, we use the Saccharomyces Genome Database R53-1-1 (April 2006) build of the \emph{S. cerevisiae} genome.

     We obtain position-specific affinity matrices (PSAMs) for a set of 26 TFs from an \textit{in vitro} microfluidics analysis of TF-DNA interactions~\cite{Fordyce2010}. This study provides PSAMs for each TF determined using the MatrixREDUCE package~\cite{Foat2006}. We convert the elements of the PSAM $w_{i\alpha}$ to EM elements using $\epsilon_{i\alpha} = -\log(w_{i\alpha})/\beta$, where $\beta =$ 1.69 (kcal/mol)$^{-1}$ at room temperature. For each of these 26 TFs, genomic sites are available in Ref.~\cite{Chen2010}. We neglect PHO4 since it does not have any binding sites above the 0.9 threshold of Ref.~\cite{Chen2010}, leaving us with 25 TFs for which both EM and a set of genomic binding sites are available. We align the binding site sequences from Ref.~\cite{Chen2010} to the corresponding EMs, choosing the alignment that produces the lowest average binding energy for the sites.


\subsection{Essentiality data}

     The Yeast Deletion Database classifies genes as essential, tested (nonessential), and unavailable, which number 1156, 6343, and 529 respectively~\cite{Winzeler1999, Giaever2002}. For each essential or tested gene, we determine all TF binding sites less than 700 bp upstream of the gene's transcription start site (on either strand), which we designate as the sites regulating that gene.
Growth rates for nonessential knockout strains are provided under YPD, YPDGE, YPG, YPE, and YPL conditions, relative to wild-type.  We choose the lowest of these growth rates to represent the fitness effect of the knockout.

     To measure the rate of nonsynonymous substitutions, we align the non-mitochondrial, non-retrotransposon ORFs taken from the Saccharomyces Genome Database R64-1-1 (February 2011) build~\cite{Cherry2012} of \emph{S.~cerevisiae} to those of \emph{S.~paradoxus} using ClustalW~\cite{Larkin2007}.  We measure the rate of nonsynonymous mutations using PAML~\cite{Yang2007}.  We ran PAML with a runMode of -2 (pairwise comparisons) and the CodonFreq parameter (background codon frequency) set to 2; we also tested CodonFreq set to zero and obtained very similar results.
We find the rate of nonsynonymous substitutions to be 0.04, and a Spearman rank correlation of $-0.16$ ($p=10^{-27}$) between growth rate of knockouts and the nonsynonymous substitution rate of the knocked-out gene. This is consistent with the results of Ref.~\cite{Zhang2005}, which found the rate of substitutions to be 0.04 and the rank correlation between growth rate and substitution rate to be $-0.19$ ($p=10^{-35}$).

To compare binding energy to evolutionary conservation, we calculate the mean Hamming distance between \emph{S.~cerevisiae} sites and corresponding sites in \emph{S.~paradoxus}~\cite{Chen2010}.
To test for significance in the difference of mean energies and Hamming distances of sites regulating essential and nonessential genes, we use a null model which assumes that the sites were randomly categorized into essential and nonessential.  We randomly choose a subset of the sites in our dataset to be ``nonessential,'' equal in size to the number of sites regulating nonessential genes as classified by the Yeast Deletion Database.  By repeating this procedure $10^7$ times, we build a probability distribution for the difference in the means of the nonessential and essential groups. The $p$-value is the probability of obtaining a difference in the means greater than the empirically measured value.


\subsection{Neutral binding site energy distributions}

We construct the neutral probability $\pi_0(\s)$ of a sequence $\s$ of length $L$ as

\begin{equation}
\pi_0(\s) = \pi_0 (\s_1) \prod_{i=2}^L \pi_0 (\s_{i-1}, \s_i),
\end{equation}

\noindent where $\pi_0 (\s_i)$ is the background probability of a nucleotide $\s_i$, and $\pi_0 (\s_{i-1}, \s_i)$ is the background probability of a dinucleotide $\s_{i-1} \s_i$.  These probabilities are determined from mono- and dinucleotide frequencies in the intergenic regions of the \emph{S. cerevisiae} genome (build R61-1-1, June 2008).  We project $\pi_0(\s)$ into energy space using Eq.~\ref{eq:seq_energy} to obtain $\pi_0(E)$, the neutral distribution of binding energies for sequences of length $L$.

     If intergenic sequences evolve under no selection with respect to their TF-binding energy, the neutral distribution of site energies should closely match the actual distribution of $L$-mer sequences obtained from intergenic regions.  Table S2, column B shows that these two distributions match very well except at the low-energy tail, which is enriched in functional binding sites.  Note that accounting for dinucleotide frequencies is important; mononucleotide frequencies alone are insufficient to reproduce the observed distribution~\cite{Djordjevic2003}.

\subsection{A model system to check the assumptions of monomorphism and steady state}

We consider a haploid asexual Wright-Fisher process~\cite{Ewens2004}. The population consists of $N_{e}=1000$ organisms, each with a single locus of $L$ nucleotides.
The new generation is created by means of a selection step and a mutation step.  In the selection step, sequences from the current population are sampled with replacement, weighted by their fitness, to construct a new population of size $N_{e}$. In the mutation step, each position in all sequences is mutated with probability $u$. For simplicity, the mutation rates between all pairs of nucleotides are the same.

     We characterize the difference between the distribution expected by our model, $\pi_\text{exp}$ (Eq.~\ref{eq:steady_state}), and the distribution observed in simulations, $\pi_\text{obs}$, using the total variation distance (TVD):

\begin{equation}
\Delta(\pi_\text{exp}, \pi_\text{obs}) = \frac{1}{2} \sum_x | \pi_\text{exp}(x) - \pi_\text{obs}(x) |.
\label{eq:TVD}
\end{equation}

\noindent The TVD ranges from zero for identical distributions to unity for completely non-overlapping distributions.  We calculate the TVD for the distributions
in energy space, where the sum in Eq.~\ref{eq:TVD} is over discrete energy bins (we bin the observed sequences by energy by dividing the range from the minimum to the maximum sequence energy for a particular EM into 100 bins of equal size). 

We begin by randomly generating the EM parameters $\epsilon_i^{\s_i}$.
Each $\epsilon_i^{\s_i}$ in the EM is sampled from a uniform distribution and then rescaled such that the distribution of all sequence energies has standard deviation of 1.0.  This is achieved by dividing all entries in the EM by a factor $\chi$:

\begin{equation} 
\chi^2 = \sum_{i = 1}^L \sum_{\alpha \in \{\mathsf{A,C,G,T}\}} \pi_0 (\alpha)  
(\epsilon_i^\alpha - \bar{\epsilon}_i )^2
\end{equation}

\noindent where $\epsilon_i^\alpha$ is the EM element for base $\alpha$ at position $i$, $L=10$ is the binding site length, $\bar{\epsilon_i} = \sum_{\alpha \in \{\mathsf{A,C,G,T}\}} \epsilon_i^\alpha$ is the average energy contribution at position $i$, and $\pi_0 (\alpha)$ is the background probability of nucleotide $\alpha$ (0.25, $\forall \alpha$ in our simulations). It can be shown that $\chi$ is the standard deviation of the random sequence energy distributution, which is approximately Gaussian~\cite{Sengupta2002}. We generate the EM once and use it in all subsequent simulations and maximum likelihood fits.

We perform the Wright-Fisher simulations in a range of mutation rates from $u = 10^{-6}$ to $u = 10^{-1}$ with a ``non-lethal'' Fermi-Dirac fitness function  (Eq.~\ref{eq:FD_fitness} with $f_0 = 0.99$, $\beta  = 1.69~(\text{kcal/mol})^{-1}$, and $\mu = -2~\text{kcal/mol}$).
We run $10^5$ simulations for each mutation rate for $100/u + 1000$ steps, enough to reach steady state. Each simulation starts from a monomorphic population with a randomly chosen sequence.  We construct the steady state distribution for each mutation rate by randomly choosing a single sequence from the final population of each simulation. Collected across all simulations, these are used to construct a distribution of sequences at each mutation rate. Additionally, we record the average final number of unique sequences at each mutation rate.

We perform another set of Wright-Fisher simulations with the same fitness function and EM as above, and $u=10^{-6}$. We run $10^5$ simulations, each starting from the same monomorphic population with a specific sequence of $E \approx 0$. At regular intervals in each simulation, we record a randomly chosen sequence from the population. Collected across all simulations, these are used to construct a distribution of sequences at each point in time.


\normalsize

\section{Acknowledgments}

A.V.M. acknowledges support from National Institutes of Health (R01 HG004708) and an Alfred P. Sloan Research Fellowship.




\bibliography{biblio}
\bibliographystyle{pnas2011}

\newpage
\section*{Figures}

\clearpage

\clearpage

\newpage
\section*{Tables}

\begin{table}[!h]
\begin{center}
\begin{tabular}{|c|p{1in}|p{1in}|p{1.5in}|p{1in}|}
\hline
TF    & $f_0$                 & $\gamma =\nu(1-f_0)$ 	& $\beta$ (in (kcal/mol)$^{-1}$)  & $E-\mu$ \\
\hline\hline
REB1  & $0.999$               & $18.3$     			& $0.801$   & $\approx 0$ \\
ROX1  & $0.992$               & $403$     			& $0.426$   & $> 0$ \\
MET32 & $0.974$               & $132$     			& $0.248$   & $> 0$ \\
RPN4  & $4.77\times 10^{-9}$  & $0.72$     			& $3.84$    & $\approx 0$ \\
MET31 & $1.85\times 10^{-10}$ & $0.547$     			& $4.63$    & $\approx 0$ \\
PDR3  & $0.789$               & $4.53\times 10^{3}$   & $0.534$   & $> 0$ \\
YAP7  & $6.01\times 10^{-6}$  & $1.26$     			& $1.13$    & $\approx 0$ \\
BAS1  & $2.09\times 10^{-3}$  & $144$     			& $0.246$   & $< 0$ \\
STB5  & $0.167$               & $168$    			& $0.301$   & $< 0$ \\
AFT1  & $3.11\times 10^{-13}$ & $0.617$     			& $16.4$    & $\approx 0$ \\
CUP9  & $0.976$               & $243$     			& $0.338$   & $> 0$ \\
MCM1  & $0.998$               & $83.8$     			& $0.25$    & $> 0$ \\
\hline
\end{tabular}
\end{center}
\caption{Summary of unconstrained Fermi-Dirac landscape fits to TF binding site data.  Columns show maximum-likelihood value of $f_0$, $\gamma = \nu(1-f_0)$, and $\beta$.  The last column shows whether most binding site energies $E$ are lower than the inferred chemical potential $\mu$, near it, or above it (see Table~S2 for details).}
\label{table:fits}
\end{table}

\begin{table}[!h]
\begin{center}
\begin{tabular}{|c|c|c|p{0.8in}|p{0.8in}|p{0.8in}|}
\hline
TF & AIC$_\text{CFD}$ $-$ AIC$_\text{UFD}$ & AIC$_\text{EXP}$ $-$ AIC$_\text{UFD}$ & $w_\text{UFD}$ & $w_\text{CFD}$ & $w_\text{EXP}$ \\
\hline\hline
REB1  &  $-2.022$  &  $35.832$  &  $0.267$   &  $0.733$   &  $4.42\times 10^{-9}$\\
ROX1  &  $-2.159$  &  $35.051$  &  $0.254$   &  $0.746$   &  $6.21\times 10^{-9}$\\
MET32 &  $-2.246$  &  $10.550$  &  $0.245$   &  $0.754$   &  $0.001$   \\
RPN4  &  $17.672$  &  $33.683$  &  $1.000$   &  $1.45\times 10^{-4}$&  $4.85\times 10^{-8}$\\
MET31 &  $2.807$   &  $11.778$  &  $0.801$   &  $0.197$   &  $0.002$   \\
PDR3  &  $-1.750$  &  $79.244$  &  $0.294$   &  $0.706$   &  $1.82\times 10^{-18}$\\
YAP7  &  $-1.988$  &  $10.783$  &  $0.270$   &  $0.729$   &  $0.001$   \\
BAS1  &  $-2.466$  &  $6.007$   &  $0.223$   &  $0.766$   &  $0.011$   \\
STB5  &  $-2.737$  &  $-7.143$  &  $0.025$   &  $0.097$   &  $0.878$   \\
AFT1  &  $-1.104$  &  $7.265$   &  $0.362$   &  $0.628$   &  $0.010$   \\
CUP9  &  $-2.284$  &  $1.689$   &  $0.219$   &  $0.687$   &  $0.094$   \\
MCM1  &  $-3.351$  &  $-0.167$  &  $0.135$   &  $0.719$   &  $0.146$   \\
\hline
\end{tabular}
\end{center}
\caption{Comparison of fitness function models. For each TF, shown are the AIC differences between the unconstrained Fermi-Dirac fit (``UFD''), the constrained Fermi-Dirac fit with $f_0 = 0.99$ (``CFD''), and the exponential fit (``EXP''). Also shown are Akaike weights $w$, which indicate the relative likelihood of each model.}
\label{table:aic}
\end{table}

\section*{Supporting Information}

\noindent\textbf{Table S1:} Full summary of tests for correlations between TF-DNA binding energies and growth rates after knockouts of genes regulated by the TF, expression levels of regulated genes, $K_A/K_S$ ratios for regulated genes,
and distances between TF sites and the TSS of the regulated gene. \\

\noindent\textbf{Table S2:} Full summary of parametric fits of fitness landscapes to TF binding site data. 

\newpage
\begin{center}
\LARGE{Supplementary Material: \\ Biophysical Fitness Landscapes \\ for Transcription Factor Binding Sites}

\vspace{0.5cm}

\large{Allan Haldane$^1$, Michael Manhart$^1$, and Alexandre V. Morozov$^{1,2}$}

\vspace{0.5cm}

	\small{\emph{$^1$ Department of Physics and Astronomy, Rutgers University, Piscataway, NJ 08854, USA}}
	
	\small{\emph{$^2$ BioMaPS Institute for Quantitative Biology, Rutgers University, Piscataway, NJ 08854, USA}}
\end{center}

\renewcommand{\thefigure}{S\arabic{figure}}
\setcounter{figure}{0}

\normalsize

\includepdf[pages=-]{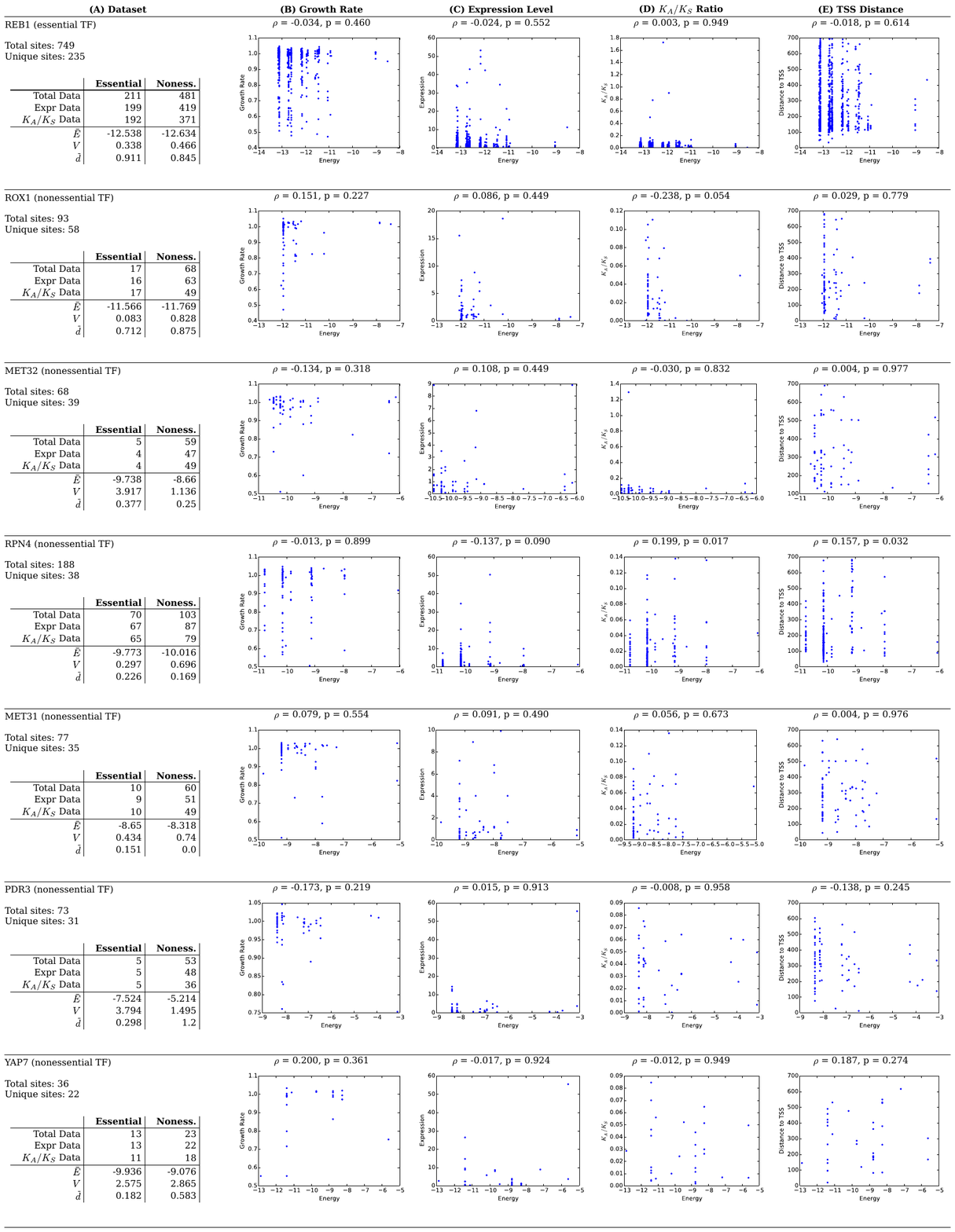}

\newpage
\begin{table}[h!]
\caption{
Full summary of tests for site-specific selection.
For 25 TFs we compute TF-DNA interaction energies (in kcal/mol) for each site. Columns from left to right:
(A)~Essentiality of the TF according to the Yeast Deletion Database; total number of binding sites for each TF; total number of sites with unique sequences. The table lists how many essential and nonessential genes are regulated by each TF, and how many of these genes have gene expression and \emph{S. paradoxus} $K_A/K_S$ ratio data. We also report the mean energy $\bar{E}$ and the variance $V$ of essential and nonessential sites, and mean Hamming distance $\bar{d}$ between \emph{S. cerevisiae} and \emph{S. paradoxus} sites regulating essential and nonessential genes.
(B)~Growth rate in strains with gene knockouts versus energy of TF binding sites regulating the knockout genes.
(C)~Gene expression versus energy of TF sites regulating the genes.
(D)~Ratio of nonsynonymous to synonymous substitutions ($K_A/K_S$) in genes versus energy of their TF regulatory sites.
(E)~Distance between each binding site and the closest transcription start site (TSS) versus the energy of the site.
For (B)--(E) we report the Spearman rank correlation $\rho$ between each property and site energy, along with the $p$-value.
}
\label{table:essentiality}
\end{table}

\includepdf[pages={1},landscape=true]{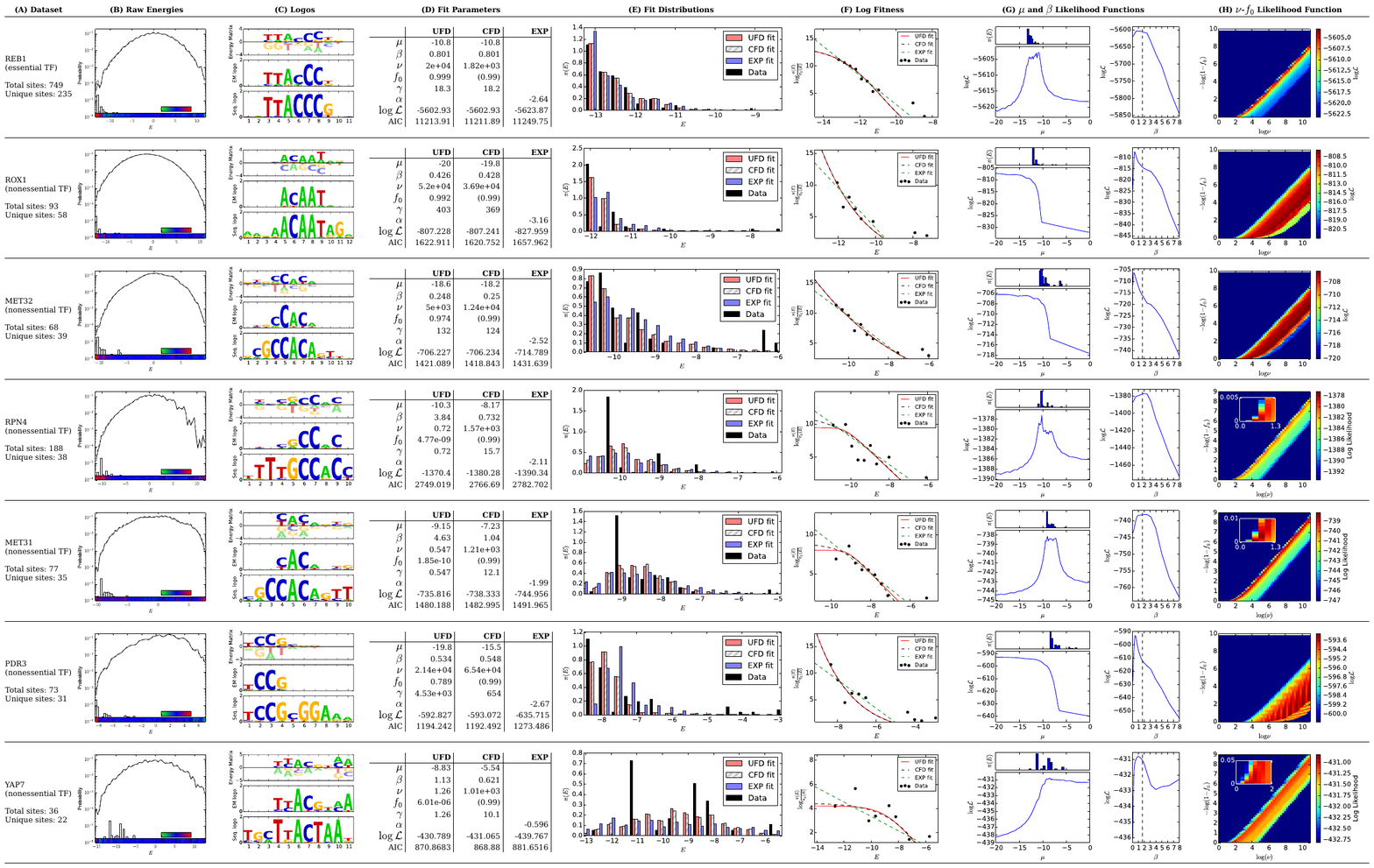}
\includepdf[pages={2},landscape=true]{FitTable.pdf}

\newpage
\begin{table}[h!]
\caption{Summary of fitness landscape fits to TF binding site data, for 12 TFs with more than $12$ unique sites. Each row corresponds to a TF, ranked in the decreasing order of the number of unique binding site sequences.  Columns, from left to right:
(A)~Summary of TF binding site data.
(B)~Same as Fig.~5B.
(C)~Same as Fig.~5A.
(D)~Fitted values of fitness landscape parameters and maximized log-likelihoods for the unconstrained fit to the Fermi-Dirac function of Eq.~7 (``UFD''), constrained fit to the Eq.~7 with $f_0 = 0.99$ (``CFD''), and fit to an exponential fitness function (``EXP'').
(E)~Same as Fig.~5D.
(F)~Same as Fig.~5C.
(G)~Left panel: Log-likelihood of the unconstrained Fermi-Dirac model as a function of the effective chemical potential $\mu$. For reference, the distribution of functional binding site energies (same as in (B)) is shown on top. Right panel: Log-likelihood as a function of the effective inverse temperature $\beta$. For reference, the inverse room temperature 1.69 (kcal/mol)$^{-1}$ is shown as the vertical dashed line. To create the log-likelihood plots, either $\mu$ or $\beta$ were held fixed while all the other parameters were re-optimized.
(H)~Heatmap of log-likelihood as a function of $\log \nu$ and $-\log(1-f_0)$ (note that $\nu(1-f_0) = \gamma =$ constant corresponds to a straight line with slope 1 in these coordinates). For likelihoods that have a maximum near $f_0 = 0$, insets show a zoomed-in view. To create the log-likelihood heatmaps, both 
$\nu$ and $f_0$ were held fixed while all the other parameters were re-
optimized. Note that in F and H, the maximum values do not always match those 
listed in D because we employ an additional round of conjugate-gradient ascent after locating the approximate maximum on the grid.
}
\end{table}

\end{document}